\documentclass[fleqn,usenatbib]{mnras}

\usepackage[english]{babel}
\usepackage{amsmath}
\usepackage{amssymb}
\usepackage{graphicx}
\usepackage{subfigure}
\usepackage[normalem]{ulem}

\usepackage{verbatim}

\usepackage{color}
\usepackage{multirow}
\usepackage{mathtools}
\usepackage{epstopdf}

\usepackage{url}
\usepackage{xcolor}

\def\nat{Nature}

\def\mnras{MNRAS}
\def\apj{ApJ}
\def\apjl{ApJL}
\def\apjs{ApJS}
\def\aap{A\&A}

\graphicspath{{figures/}{../figures/}}

\usepackage{newtxtext,newtxmath}

\usepackage[T1]{fontenc}

\DeclareRobustCommand{\VAN}[3]{#2}
\let\VANthebibliography\thebibliography
\def\thebibliography{\DeclareRobustCommand{\VAN}[3]{##3}\VANthebibliography}

\usepackage{graphicx}	
\usepackage{amsmath}	
\usepackage{amssymb}	

\usepackage{hyperref} 
\usepackage{natbib}

\newcommand{\msun}{\ensuremath{\rm{M_{\odot}}}}
\newcommand{\e}{\ensuremath{ \epsilon_{\star} }}

\title[UFDs: the minimum mass of the first stars]{Ultra-faint dwarf galaxies: \\ unveiling the minimum mass of the first stars}

\author[M. Rossi et al.]{
Martina Rossi$^{1,2}$\thanks{E-mail:martina.rossi@unifi.it},
Stefania Salvadori$^{1,2}$, and
{\'A}sa Sk{\'u}lad{\'o}ttir$^{1,2}$
\\
$^{1}$Dipartimento di Fisica e Astrofisica, Univerisitá degli Studi di Firenze, via G. Sansone 1,Sesto Fiorentino,Italy\\
$^{2}$INAF/Osservatorio Astrofisico di Arcetri, Largo E. Fermi 5, Firenze, Italy
}

\date{Accepted  2021 March 16, Received 2021 March 16, in original form 2020 November 18}

\pubyear{2020}

\begin{document}
\label{firstpage}
\pagerange{\pageref{firstpage}--\pageref{lastpage}}
\maketitle

\begin{abstract}
The non-detection of zero-metallicity stars in ultra-faint dwarf galaxies (UFDs) can be used to constrain the Initial Mass Function (IMF) of the first (PopIII) stars by means of a statistical comparison between available data and predictions from chemical evolution models.
To this end we develop a model that follows the formation of isolated UFDs, calibrated to best reproduce the available data for the best studied system: Bo{\"o}tes~I. Our statistical approach shows that UFDs are the best suitable systems to study the implications of the persisting non-detection of zero-metallicity stars on the PopIII IMF, i.e. its shape, the minimum mass~($m_{min}$), and the characteristic mass~($m_{ch}$). 
We show that accounting for the incomplete sampling of the IMF is essential to compute the expected number of long-lived PopIII stars in inefficiently star-forming UFDs. By simulating the Color Magnitude Diagram of Bo{\"o}tes~I, and thus take into account the mass-range of the observed stars, we can obtain even tighter constrains on $m_{min}$. By exploiting the 96 stars  with measured metallicities ($\rm i < 19$) in the UFDs represented by our model, we demonstrate that: $m_{ch} > 1 \msun$ or $m_{min} > 0.8 \msun$ at $99\%$ confidence level. This means that a present day IMF for PopIII stars is excluded by our model, and a top-heavy PopIII IMF is strongly favoured. We can limit $m_{min} > 0.8 \msun$ independent of the PopIII IMF shape by targeting the four UFDs  Bo{\"o}tes I, Hercules, Leo IV and Eridanus II with future generation instruments, such as ELT/MOSAIC ($\rm i < 25$), which can provide samples of >10\,000 stars.

\end{abstract}




\section{Introduction}
The first stars are expected to have played a crucial role in the evolution of the primordial Universe since they represent the first sources of ionising photons, dust and chemical elements heavier than helium, i.e. metals. The amount of ionizing photons, dust, and different chemical species produced by the first stars strongly depends on their mass. Thus, understanding the Initial Mass Function (IMF) of the first stars 
is a fundamental problem in Cosmology. Yet, this function is very difficult to infer, both from an observational and a theoretical prospective. \\
First stars, also know as Population III (PopIII) stars, are predicted to form at $ z\sim15-30$, in so-called \textit{minihaloes} with masses $\rm M_h \approx 10^6 \: \msun$ and virial temperature $ \rm T_{vir} \le 10^4$ K (e.g. \cite{bromm02}, \cite{yoshida03}). Since they form out of gas of primordial composition they are expected to be completely metal-free. These metal-free stars should produce key emission lines that might allow us to easily identify their host galaxies (e.g. \cite{Schaerer02}). However, since PopIII stars predominantly formed in poorly star forming minihaloes, they are likely too faint to be detected, even with new-generation telescopes (e.g. \cite{Wise12}). 
Therefore we have to deal with a lack of direct observations of PopIII stars to infer their mass distribution.

\noindent Our current knowledge of the mass properties of PopIII stars derives on one hand from computer simulations, and on the other from Stellar Archaeology. The latter indirectly provides information about the first stars by exploiting the stellar chemical abundances measured in the oldest and most metal-poor components of our Milky Way and its dwarf galaxies satellites  (e.g. \citet{mcwilliam98}, \citet{freeman02}, \citet{gratton04}, \citet{venn04}, \citet{beers05}, \citet{tolstoy09}, \citet{asa15}, \citet{Frebel15}, \citet{ji16}, \citet{bonifacio18}, \citet{li18}, \citet{Starkenburg18}, \citet{aguado19}, \citet{Bonifacio19}, \citet{sestito19}, \citet{reichert20}, \citet{chiti21}).  Various different works (e.g. \citet{SS15}, \citet{Hartwig15}, \citet{debennessuti17}, \cite{magg17}, \citet{ishigaki18}) have interpreted these results as indirect evidence of the of the existence of massive PopIII stars ($[10-60] \: \msun$), as well the proof of existence of pair-instability supernovae, $\rm m_{PopIII} = [140-260] \: \msun$ (\citet{aoki14}, \citet{SS19}, \citet{Tarumi20}).\\

\vspace{0.1in}

\noindent In parallel, hydrodynamic simulations of PopIII stars formation predict that the first stars are more massive than stars formed today, with characteristic masses of $[10-20] \:  \rm{M_{\odot}}$ and the maximum mass possibly extending up to $1000 \: \rm{M_{\odot}}$ (\cite{omukai01}, \cite{abel02}, \cite{bromm02}, \cite{tan04}, \cite{yoshida06}, \cite{oshea07}, \cite{hosokawa11}, \cite{Bromm13}, \cite{Hirano14}). However, 3D simulations that study the cooling of primordial gas reveal that proto-stellar gas clouds can experience strong fragmentation and PopIII stars could have masses lower than $1 \:\msun$ (e.g. \cite{machida08}, \cite{turk09}, \cite{smith10}, \cite{clark11}, \cite{greif11}, \cite{stacy13}, \cite{dopcke13}, \cite{stacybromm14}, \cite{stacy16}, \cite{susa19}, \citet{wollenberg20}). Yet, we have to stress that these high-resolution simulations follow the evolution of the proto-star for a very short time compared to the time-scale of star formation. Hence we do not know if these fragments will eventually merge into the central object (e.g. \cite{HiranoBromm2017}) or can be thrown out and form long-lived first stars that may have survived until today.\\

\noindent As a result of the unpredictable and stochastic star formation process, the mass distribution of the first stars is expected to cover a wide range of masses, PopIII $m_{\star} = [0.1-1000]\msun$, possibly reaching down into the sub-stellar regime. However, if PopIII stars with mass below $0.8\msun$ were able to form they should be alive today, in particular in the oldest and least metal-poor systems of the Local Group, where stars can be observed individually.\\ 

\noindent Despite long searches, zero-metallicity stars have never been found among the ancient stars in our Milky Way and its dwarf galaxy satellites. The apparent absence of long-lived PopIII stars can be attributed to two main reasons: the first is that PopIII stars are so rare and challenging to find that we haven't caught them yet; the second is that PopIII stars were more massive than present-day stars and hence did not survive until today, i.e. with minimum mass $m_{min} > 0.8\msun$ (e.g. \cite{oey}, \cite{SS07}, \cite{tum07}, \cite{SS10}). Non-detection of long-lived Pop III stars can thus be an instrument to constrain the lower mass limit of the PopIII IMF. Using a statistically approach, \cite{Hartwig15} have, however, recently shown that is not so straightforward to exclude the existence of any PopIII stars survivors and set PopIII $m_{min}\ge 0.8\msun$. Indeed, a large observed stellar sample is required: $\sim 10^7$ stars for the Milky Way's halo and $\sim 10^{10}$ for the Bulge. However these numbers are much larger than what can be expected in the largest spectroscopic surveys in the next decade (e.g. 4MOST: \citet{dejong4MOST}).

\noindent At the moment, therefore, we have only a few observational-driven constraints on the IMF of the PopIII stars and we are not even able to exclude that it was equal to present-day stellar IMF. This brings us to the question: How can we further constrain the lower mass-end of the PopIII IMF? \\ 

\noindent Among all environments hosting ancient stars, ultra-faint dwarf  (UFD) galaxies ($\rm L_{\rm bol} < 10^5$ $\rm L_{\odot}$) are the best places to look for the most metal-poor stars. This has been shown both theoretically (e.g. \cite{SS09},  \cite{SS10}, \cite{magg17}) and observationally (e.g. \cite{ishigaki14}, \cite{webster15}, \cite{Frebel16}, \cite{Pakahmow19}). By modelling early star formation, \cite{magg17} estimate the numbers of surviving PopIII stars in the Milky Way and its dwarf companions. Their results show that UFD satellites are the best systems to look for these stellar relics, since they typically contain a larger fraction of Pop III star survivors, up to $0.74\%$ of the total stellar mass. Furthermore, many theoretical studies have shown that UFD galaxies have the simplest assembly histories with no major merger events. Unlike essentially all larger systems, they underwent little to no further evolution and have survived to the present day as pristine relics from the early Universe. For this reason they are generally considered the $``fossils"$ of the minihaloes that hosted the first stars (e.g. \cite{Ricotti2005}, \cite{Bovill09Ricotti}, \cite{SS09}, \cite{BH2015}, \cite{SS15}).  \\

\noindent Observationally, UFD galaxies are the most common dwarf galaxies in the Local Group, representing more than $50\%$ of the total number of dwarf satellites (\cite{Mccon12}). They are the oldest, most dark matter-dominated, most metal-poor, least luminous, and least chemically evolved stellar systems known (\cite{Simon19}). Since they have the smallest dark matter haloes they are more sensitive to feedback processes, representing the extreme limit of the galaxy formation processes. Most UFD galaxies formed more than $75 \%$ of their stars in the first Gyr of evolution and hence have old stellar populations, $\gtrsim10$ Gyr (e.g. \cite{Brown14}). Recent Hubble Space Telescope (HST) observations of Eridanus~II revealed that in this distant UFD galaxy $>75 \%$ of the stars have ages $>13$~Gyr and possibly formed in a very short burst, lasting $<200$~Myr (\citet{Eridanus}). Finally, UFDs are the galaxies that contain the highest fraction of extremely metal-poor stars, $\rm[Fe/H] < -3$ (e.g.~\cite{kirby13}, \cite{SS15}).\\

\noindent In this work we investigate the frequency of PopIII star survivors in UDF galaxies to constrain the minimum mass, $m_{min}$, and the characteristic mass, $m_{ch}$, of the first stars. To this end we develop a semi-analytical model that follows the formation and evolution of an isolated UFD galaxy. The ideal target for our study is Bo{\"o}tes~I since this is the UFD galaxy with the best observational data (e.g. \cite{Frebel10}, \cite{norris10}, \cite{Lai11}, \cite{gilmore13}, \cite{Frebel16}), and it has also been studied extensively from a theoretical point of view (e.g. \cite{vincenzo14}, \cite{romano14}). The color-magnitude diagram of Bo{\"o}tes I reveals that its stellar population is old and metal-poor, with a mean age $\sim (13.3 \pm 0.3)~ \rm Gyr $ (\cite{Brown14}). Given the higher luminosity  of Bo{\"o}tes I ($\rm L_{\star} \sim 10^{4.5} \rm L_{\odot}$) with respect to other UFDs, this is also the only UFD galaxy for which we can derive an almost complete Metallicity Distribution Function (MDF; \cite{Lai11}, \cite{norris10}, \citet{gilmore13}).\\

\noindent The article is structured as follows:  Sec.\ref{modeldescriprtion} describes the main features of the semi-analytic model whose data-calibration process is presented in Sec.\ref{modelcal}.  In Sec.\ref{samplingdescription} it is explained why it is fundamental to account for the incomplete sampling 
of the stellar IMF in the modelling of poorly star-forming UFDs, in particular to compute the expected fraction of PopIII stars survivors. The observationally driven constraints for the minimum mass of PopIII stars are presented in Sec.\ref{impactIMF}, while in Sec.\ref{strong} we demonstrate how to get stronger constraints by computing the synthetic colour magnitude diagram, and thus compare model predictions with truly observed stars. Finally the summary and the discussion of results are in Sec.\ref{summary}
\newpage
\section{Model Description}
\label{modeldescriprtion}
Our semi-analytic model follows the star formation and the chemical enrichment history of an UFD galaxy from the epoch of its virialization $(z = z_{\rm vir})$ until present day $(z =0)$. 
We model the evolution of present-day UFD galaxies by assuming that they experienced a $``$quiet$"$ assembling history, i.e. without major merger events. The evolution in isolation is a good approximation for these tiny systems, as has been shown by cosmological models and simulations for the Local Group formation (e.g. \cite{SS09}, \cite{SS15}, \cite{safarz18}), as well as recent observations (\citet{Eridanus}).\\

\begin{itemize}
\item{\bf{Initial Conditions:}}

\begin{figure} 
	\includegraphics[width=\columnwidth]{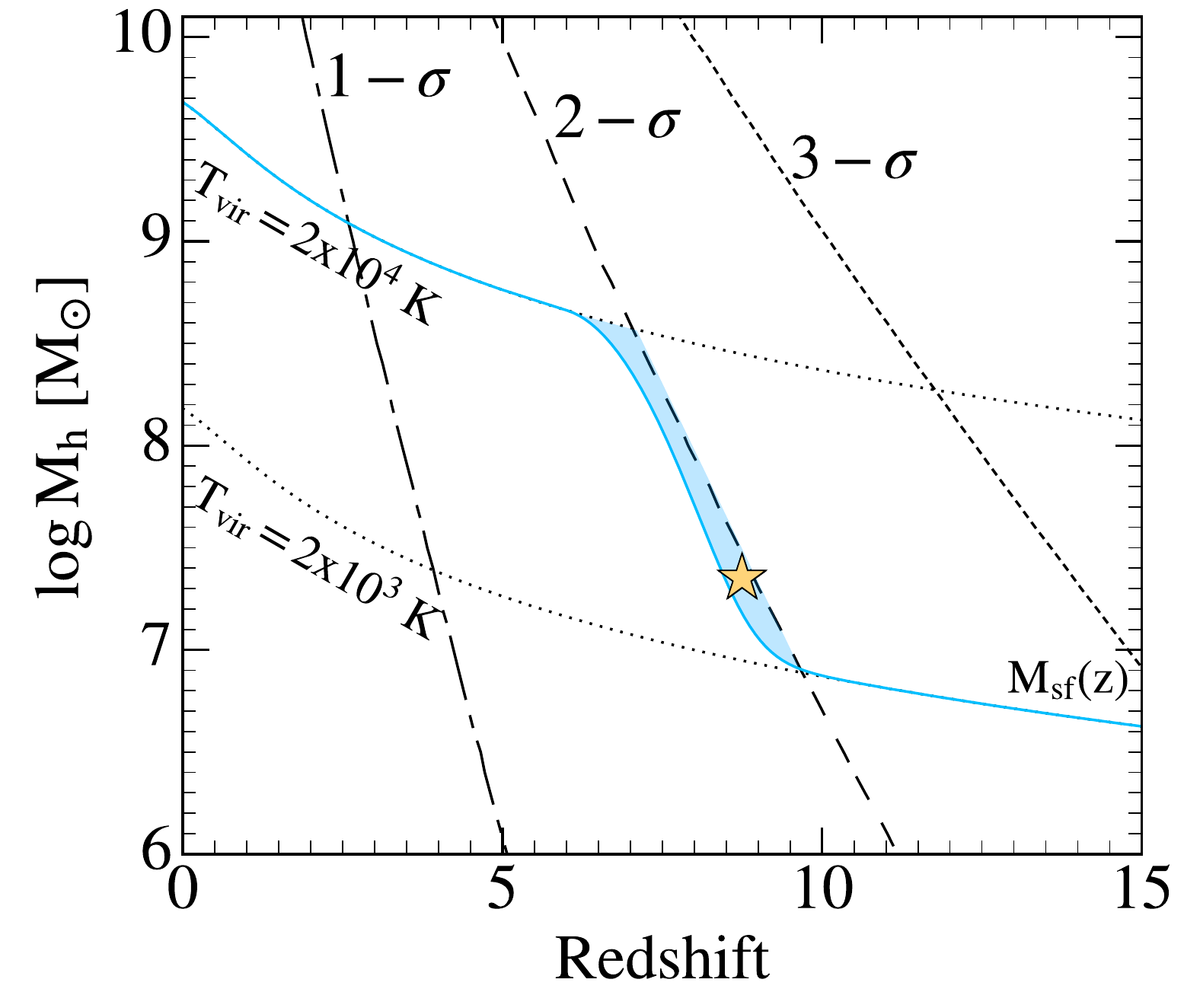}
    \caption{ Evolution of the minimum halo mass to form stars, 
    $\rm{M_{sf}}(z)$, (solid blue line), of the halo mass corresponding to
    1-, 2-, 3-$\sigma$ fluctuations of the density field (long-short, long, and short dashed lines), and of the halo mass with virial temperature
    $\rm T_{vir} = 2 \cdot 10^3 K$ and $\rm T_{vir} = 10^4 K$ (dotted lines) adapted from \citet{SS09}. The light blue shaded area shows the
    condition to form satellite dwarf galaxies, $\rm {M_{sf}(z)} < M_h < M_{2\sigma}$. The yellow star represents our choice for the initial conditions: $\rm M_{h} = 10^{7.35}\: \rm M_{\odot}$ and $z_{vir}=8.7$.
    \label{mh}} 
\end{figure}
\noindent The nature of UFD galaxies is in ongoing debate. However, their observed properties suggest that they could be the living relics of the first star-forming minihaloes that hosted PopIII stars. This scenario has been investigated, for example, by \cite{SS09} that explored the origin of UDF galaxies using a cosmological model for the formation of the Milky Way and its dwarf satellites, including UFD galaxies. These authors selected Milky Way's satellite candidates among star-forming haloes that correspond to $< 2-\sigma$ density fluctuations\footnote{The quantity $\sigma(M, z)$ represents the linear rms density fluctuation smoothed with a top-hat filter of mass M at redshift $z$.} because these objects are the most likely to become satellites (\cite{diemand05}). By comparing their model results with observations, they concluded that UFD galaxies are associated to minihaloes formed prior to the end of reionization, $z_{\rm rei} > 6$, which have not experienced mergers, i.e. they evolved in isolation (see also \cite{SS15}). Following this approach we select the dark matter haloes that correspond to $< 2-\sigma$ fluctuations of the density field\footnote{We adopt $\Lambda$ cold dark matter ($\Lambda$CDM) with $h = 0.669$, $\Omega_{\rm m}=0.3103$, $\Omega_{\Lambda}=0.6897$, $\Omega_{\rm b} h^2 = 0.02234$, $\sigma_8 = 0.8083 $, $n= 0.9671$, according to the latest Plank results (\cite{plank18})} and that are able to form stars. The latter condition is described by the minimum halo mass required to allow star-formation, $\rm M_{sf}(z)$, see Fig.\ref{mh}, whose evolution accounts for the increasing Lyman Werner and ionizing radiation (see Sec.2.1 \cite{SS09}). In Fig.\ref{mh} our UFD galaxies candidates are those with dark matter halo mass $\rm {M_{sf}(z)} < M_h < M_{2\sigma}$ (indicated with the shaded light blue area). Following the results of \cite{SS15} we individuated the progenitor of Bo{\"o}tes~I in the halo with dark matter halo $\rm M_h = 10^{7.35}$ $\msun$ that virializes at $z_{vir}= 8.7$ (yellow star in Fig.\ref{mh}) and adopt these values as initial condition of our model.\\

\noindent We assume that at the epoch of virialization the dark matter halo contains a total amount of gas equal to $\Omega_{\rm{b}}/\Omega_{\rm m} \: \rm  M_h$. 
The star formation becomes possible when the gas begins to infall in the central part of the halo and to cool down. Considering the redshift of virialization, the infalling gas can be assumed to be metal-free (e.g. \cite{SF12}).\\
\item{\bf{Infall rate:} }\\
Following \cite{SS08,SS15} we exploit the results of simulations presented in \cite{Ker15} to compute the infall rate as:
\begin{equation}
 \frac{dM_{inf}}{dt} = A { \left( \frac{t}{t_{inf}} \right)}^2 \exp { \left(- \frac{t}{t_{inf}} \right)}
\label{Minfall}
\end{equation}
where A is the normalization constant, which is set to be $\rm A=2 M_{h} \frac{\Omega_{\rm {b}}}{\Omega_{\rm{ m}}}\frac{1}{ t_{inf}} $ so that for $ t \longmapsto \infty$ the accreted gas reaches the initial value, $  M_{inf}(\infty)= \rm \Omega_{\rm{b}}/\Omega_{\rm {m}} \: M_h$. In addition, $t$ is the time since virialization  and $t_{inf}$ is the time-scale on which the gas cools, which is a free parameter of the model.\\
\item{\bf{Star formation rate:}}\\
It has been assumed that star formation occurs in a free-fall time $t_{ff} = (3\pi/ 32G\rho(z))^{1/2}$, where G is the gravitational constant and $\rho$ the total mass density inside the halo. Stars are assumed to form in a single burst at each time step, $dt$, and the star formation rate, $\Psi$, is given by:
\begin{equation}
 \Psi =\frac{dM_{\star}}{dt}= \epsilon_{\star} \frac{M_{gas}}{t_{ff}(z)}  
\label{stelle}
 \end{equation}
 where $\rm M_{\star}$ is the total stellar mass formed, $\rm M_{gas}$ is the mass of the cold gas, and $\e$ is the star formation efficiency, a second free parameter of our model.\\
\item{\bf{PopIII and PopII/I stars:}}\\
\label{pop3andpop21}
The metallicity of the cold gas of the interstellar medium (ISM), $\rm Z_{gas} = M_{Z}^{ISM} / M_{gas}$, 
determines whether the stars forming are PopIII or PopII/I stars. PopIII stars only form if the metallicity of the gas is lower than the critical metallicity, $\rm Z_{cr} = 10^{-4.5} Z_{\odot} $, (\cite{debennessuti17}), instead PopII/I stars form if $\rm Z_{gas} > Z_{cr} $. At each time step we compute the stellar mass formed, and we assume that the star formation occurs only if a minimum value is reached, i.e. if $\rm M_{\star}(t_{i}) \geq \rm M_{\star}^{min}=100 \: \rm{M_{\odot}} $ (\cite{klessen11}).
\newpage
\item{\bf{Stellar initial mass function:}}\\
\label{stellarinitialmassfunction}
The stellar mass formed at each time step is distributed according to an IMF. As a first approximation we assume that the stellar mass formed is enough to fully populate the IMF. In other words, the total mass of stars formed at each time-step is distributed throughout the overall stellar mass range. As stellar IMF for PopII/I stars we choose a Larson-type (\cite{Lars98}): \\
	\begin{equation}
	\phi{(m_{\star})} = \frac{dN}{ dm_{\star}} \propto  {m_{\star}^{-2.35}} {exp{\left(-{m_{ch} \over m_{\star}} \right)}}
	\label{larson}
	\end{equation} \\
where the stellar mass $m_{\star}$ varies in the mass range $[0.1 - 100]  \rm{M_{\odot}}$ and $m_{ch} = 0.35\rm{M_{\odot}} $ is the characteristic mass.
Since the PopIII IMF is unknown, we initially assume it to be equal to the IMF of present-day stars (\ref{larson}, see Fig.\ref{imfs}) and we explore the impact of different minimum mass of PopIII stars, $m_{min}$. Then (Sec.~\ref{impactIMF}) we vary both the maximum mass of PopIII stars, $m_{max}$, and the PopIII IMF shape by assuming three different $m_{ch}$ along with a Flat IMF. All our explored PopIII IMFs are summarized in Fig.\ref{imfs}. \\
\begin{figure}
	\includegraphics[width=\columnwidth]{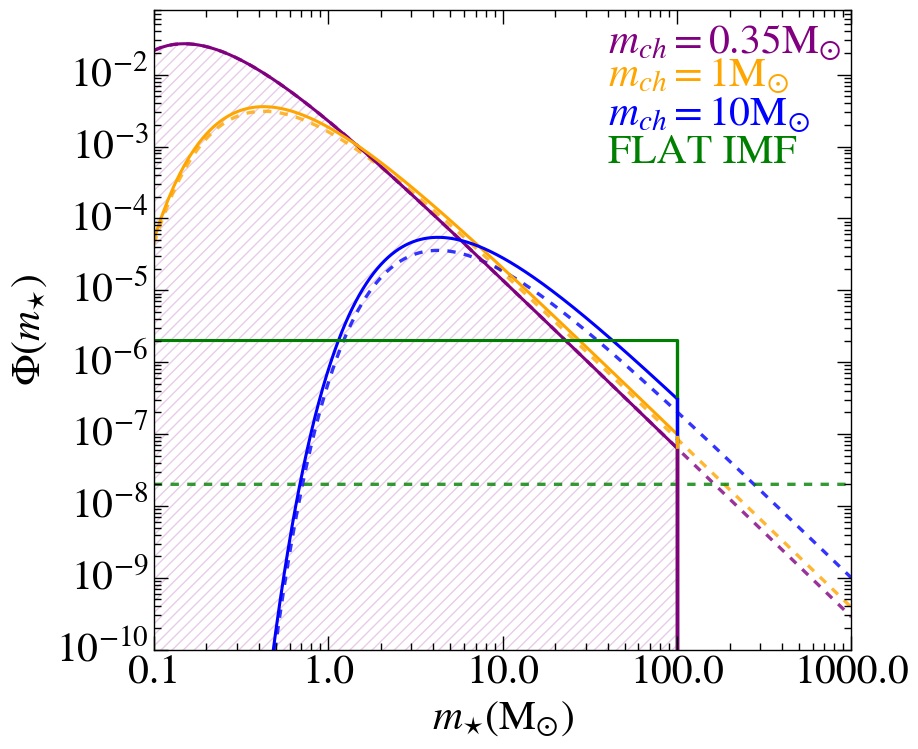}
    \caption{Normalized PopIII IMFs explored in our model (see labels). Solid lines represent IMFs in the mass range  $[0.1 - 100]  \: \rm{M_{\odot}}$, while dashed lines have maximum mass, $m_{max} = 1000 \msun$. The hatched area highlights the IMF of present-day stars \citep{Lars98}.
    }
    \label{imfs}
    \end{figure}


\item{\bf{Stellar evolution:}}	\\
\label{evolution}
The model accounts for the finite stellar lifetimes, i.e. for the evolution of stars with different masses. We use the relation derived by \cite{Raiteri96} in which the stellar lifetime depends not only on the stellar mass but also on the stellar metallicity, $t_{\star}(m_{\star}, \rm Z_{\star})$, which is settled by the ISM out of which stars formed. All stars whose lifetime is greater than the difference between the age of the Universe at $  z=0 $ ($\sim13.8 $~Gyr) and the age of the Universe when they formed, $t_{\star} > (t_{z=0}- t_{z_{form}})$, can survive until today (see details in \cite{SS08}). In particular, according to the adopted lifetime-mass-metallicity relations, all stars with $m_{\star} \le 0.8 \msun$ have a lifetime greater than the age of the Universe and can therefore survive until today independent on their formation redshift. 
\begin{enumerate}
\item[-] {\it{Returned fraction and metal yields}}:\\
Stars return gas into the ISM, via supernova (SN) explosions or via stellar winds in the Asymptotic Giant Branch (AGB) phase. This gas becomes thus newly available for star formation. 
We compute the rate at which the gas is returned into the ISM as:\\
\begin{equation}
\label{returned}
 \frac{dR(t)}{dt} = \int_{m_{turn}(z)}^{100 \msun} (m_{\star}-w_{m}(m_{\star})) \Phi(m_{\star}) \Psi(t - t_{{\star}}) dm_{\star}
\end{equation}\\
where $\ w_{m}(m_{\star})$ is the remnant mass of a star with initial mass $\ m_{\star} $, $ \Psi$ the star formation rate, and $ t_{{\star}}  = t_{\star}(m_{\star}, \rm Z_{\star})$ is the stellar lifetime.\\ 

\noindent The gas that is returned to the ISM carries with it the products of nuclear burning, i.e metals ($\rm Z$). The rate of heavy elements newly synthesised inside stars and re-ejected into ISM is:
\begin{equation}
\label{metalyields}
\begin{split}
 &\frac{dY_{Z}(t)}{dt} = \int_{m_{turn}(z)}^{100 \msun} (m_{\star} -w_{m}(m_{\star})-  m_{Z}(m_{\star},  {Z_{\star}})){{Z}}(t- t_{\star} )+\\
& \qquad \qquad \qquad \qquad  m_{Z}(m_{\star},{Z_{\star}})) \Phi(m_{\star}) \Psi(t - t_{\star}) dm_{\star}
\end{split}
\end{equation}\\
where $ m_{Z}(m_{\star}, Z_{\star})$ is the mass of  heavy elements produced by a star with initial mass $  m_{\star}$ and metallicity $\rm Z_{\star}$, and $ Z(t- t_{\star}) $ is the abundance of metals in the ISM at the time $(t- t_{\star})$. In our model we follow the evolution of the total amount of heavy elements ($\rm Z $) and of iron ($\rm Fe $). The values used in our model to compute the metal yields and returned fraction of gas are those derived by \cite{Woo95} for both PopII/I and PopIII stars with masses $<100\msun$, while for PopIII stars with $\rm 140\msun \leq m_{PopIII} \leq 260\msun$ we adopt the yields by \citet{heger02}.
\end{enumerate}
\item{\bf{Mechanical feedback:}}\\
Stars with masses $m_{\star} = [8 -40]\rm{M_{\odot}}$ end their life as SN. SNe explosions power a blast wave which, if sufficiently energetic, may overcome the gravitational pull of the host halo leading to expulsion of gas and metals into the intergalactic medium (IGM). We assume the mass of gas ejected into the IGM due to SN explosions to be regulated by the equation: 
	\begin{equation}
 	\frac{1}{2} M_{ej} v_{esc}^{2} = E_{SN} 
	\label{esn}
	\end{equation}
	\noindent where $\rm E_{SN}   = \epsilon_{w}\ N_{SN} \left \langle E_{SN} \right \rangle $ is the total kinetic energy injected by SN-driven winds, $\rm N_{SN}$ the number of SNe, $\rm \left \langle E_{SN} \right \rangle $ the average SN explosion energy, and $\epsilon_{w}$ is a free parameter, which controls the conversion efficiency of the explosion energy into kinetic form. Assuming the IMF described by \ref{larson}, we get that the number of SNe per unit of stellar mass formed $ \eta_{SN} $ is: 
\begin{equation}
   \eta_{SN} = \frac{ \int_{ 8 \msun }^{40 \msun} m_{\star} \: \Phi(m_{\star}) dm_{\star}
}{\int_{0.1 \msun }^{100 \msun}  \: \Phi(m_{\star}) dm_{\star}} \approx 10^{-2} 
\end{equation}
 For the average value of the SN energy we assume  $\rm \left \langle E_{SN}\right \rangle \sim 10^{51}~ \rm erg$ for PopII SNe, which is a typical value for a normal core-collapse SNe. For   
 for massive PopIII stars ($[140 - 260] \msun$), i.e. Pair Instability SNe (PISN), we use the mass-energy relation of \citet{heger02}, in which the energy increases with the mass of PISN. For less massive PopIII stars ($[8 - 40] \msun$) we assume the same energy of PopII SNe. \\

\noindent Following \cite{SS08} we differentiate Eq.\ref{esn} to get the gas ejection rate:
	\begin{equation}
	\label{mej} 
	\frac{dM_{ej}}{dt} =\frac{2\ \epsilon_{w} \left \langle E_{SN} \right \rangle }{v_{esc}^{2}} \: \frac{dN_{SN}}{dt} \approx \frac{2\ \epsilon_{w} \left \langle E_{SN} \right \rangle }{v_{esc}^{2}} \: 10^{-2} \: \frac{dM_{\star}}{dt}		\end{equation}
and compute the escape velocity of the gas as a function of the halo virial radius, $v_{esc}^{2} = \frac{G M}{r_{vir}}$ (\cite{Bark01}).

\noindent In our model we assume that the metallicity of the ejected gas is the same as that of the ISM. This is an approximation, we are assuming that there is a complete mixing between the gas inside the halo and the heavy elements injected into the ISM by SN explosions.\\

\noindent A limitation of this study is that only core-collapse SNe  ($m_{\star}=[8-40]\msun$) have been considered for PopII/I stars and we have not included the contribution of SN~Ia. The choice was guided by the fact that the evolutionary timescales associated with SN~Ia are typically very long, $>1$Gyr, compared to the evolution time-scale of our galaxy ($\sim500$Myr). Although there are studies that have shown that the evolutionary scale times of SN~Ia can be shorter (e.g. \cite{matteucci06}), cosmological models for UFDs formation that also include prompt formation of SNIa show that their influence on the chemical evolution is extremely limited, and is only relevant at the highest $\rm [Fe/H]$ (\cite{SS15}). In fact, a detailed chemical abundance study of Bo{\"o}tes~I did not reveal significant SNIa contribution (\cite{gilmore13}). Therefore their contribution will only be considered in future work. Finally, we considered the evolution of an isolated UFD galaxy, however to obtain a more accurate and realistic model the cosmological context should be considered. \\
\item{\bf{Galaxy evolution:}} \\
To follow the evolution of an UDF galaxy we solve Eq.~\ref{stelle} along with the subsequent system of differential equations:
 \begin{equation}
\frac{dM_{gas}}{dt}= -\Psi + \frac{dR}{dt} + \frac{dM_{inf}}{dt} - \frac{dM_{ej}}{dt} 
\label{mgas}
\end{equation} 
 \begin{equation}
\frac{dM_{Z}}{dt}= -Z^{ISM} \ \Psi +\frac{dY_{Z}}{dt} + Z^{inf}\: \frac{dM_{inf}}{dt} - Z^{ISM}\: \frac{dM_{ej}}{dt}
\label{mz}
\end{equation}\\
where $ \rm M_{gas}$ in Eq.\ref{mgas} is the mass of cold gas inside the halo, which increases due to the infall rate (Eq.\ref{Minfall}) and to the amount of gas released by massive and AGB stars (Eq.\ref{returned}), and it decreases due to star formation (Eq.\ref{stelle}) and gas ejection caused by SN explosions (Eq.\ref{mej}). The second equation, Eq.\ref{mz}, describes the variation of the total mass of metals ($\rm M_{Z}$) in gas where  $\rm Z^{ISM} $, $\rm Z^{inf} $ and $\rm Z^{ej} $ are respectively the metallicity in the ISM, in the infalling gas, and in the ejected gas. In our model we assume $\rm Z^{inf} = 0 $, i.e. the infalling gas is considered metal (and iron free) and $\rm Z^{ej} = Z^{ISM}$ (see previous paragraph). Thus, we can write the fourth equation that describes the evolution of the iron mass in the ISM as:
 \begin{equation}
 \frac{dM_{Fe}}{dt}= -\frac{M_{Fe}^{ISM}}{M_{gas}}\ \Psi+\frac{dY_{Fe}}{dt} - \frac{M_{Fe}^{ISM}}{M_{gas}} \: \frac{dM_{ej}}{dt}
\label{mfe}
\end{equation}\\
where $\rm M_{Fe}^{ISM}$ is the total iron mass in the ISM.
\end{itemize}
\subsection{Numerical Code}
\label{numerical}
To follow the evolution of the UFD galaxy through cosmic time we solve the set of the differential equations (\ref{stelle}-\ref{mfe}), integrating them step by step. To this end we build a grid of cosmic time values from $ t_{form} = t( z_{vir}=8.7)$ to $t(z=0)$ with a constant time step 
$ \Delta t =( t_{i+1} - t_{i}) = 1 \rm Myr$. This time step is adequate to describe the formation and evolution of the stars and ensures that $\Delta t < t_{ff}(z)$, thus allowing us to accurately follow the star formation. Furthermore, it is smaller than the lifetime of the most massive stars contributing to chemical enrichment, $ \Delta t <  t_{\star}(260 \: \rm{M_{\odot}}) \sim 4$ Myr. \\

\noindent To make comparison with observations, we derive the cumulative star formation history ($\textit{Cumulative SFH} $) that is defined as the ratio between the stellar mass formed until $  t_{i} = t(z_{i})$ and the {\it total} stellar mass formed, i.e. until $z=0$:
 \begin{equation}
Cumulative \: SFH(i) =  \frac{\sum_{t_{i}=t_{vir}}^ {t(z_{i})} M_{\star}(t_{i})}{\sum_{t_{i}=t_{vir}}^ {t(z=0)} M_{\star}(t_{i})  } 
\end{equation}
At each time step $t_{i}$, the evolution of the mass of gas, $\rm M_{gas}$$(t_{i})$, metals, $\rm M_{Z}$$(t_{i})$, and iron, $\rm M_{Fe}$$(t_{i})$ are computed. These quantities allow us to calculate the metallicity of the stars, $\rm Z_{\star}=M_{Z}/ M_{gas}$, and their iron abundance $\rm [Fe/H]$:
\begin{equation}
{\rm[Fe/H]}= \log \left( \frac{N_{Fe}}{N_{H}}\right)  -  \log \left( \frac{N_{Fe}}{N_{H}} \right)_{\odot}  \approx \log \left( \frac{M_{Fe}}{M_{gas}}\right)  -  \log \left( \frac{M_{Fe}}{M_{H}}\right)_{\odot} 
\label{iron}
\end{equation} 
where $\rm N_{Fe}$ ($\rm N_{H}$) is the number of iron (hydrogen) atoms and we used $ \rm M_{Fe} = \rm N_{Fe} \cdot m_{Fe}$ and $\rm M_{gas} \approx M_{H} = \rm  N_{H} \cdot m_{H}$. For the solar values we assume $\frac{M_{Fe}}{M_{H}}_{\odot} =\: 1.27 \cdot 10^{-3}$ and $Z_{\odot} = 0.02$ (\cite{anders}).\\

\noindent  Finally, at each time step we keep track of the number, mass and iron abundance of stars that survive until $z=0$, in order to determine the total stellar mass of surviving stars and their MDF. From the MDF we derive the average stellar iron abundance $\left \langle \rm [Fe/H] \right \rangle$, while we get the galaxy luminosity, $\rm L_{\star}$, from the total stellar mass of surviving stars by assuming $\rm M_{\star}/ L_{\star} \approx 1 $ (e.g. \cite{SS08}).

\section{Model Calibration}
\label{modelcal}

\begin{figure}
	\includegraphics[width=\columnwidth]{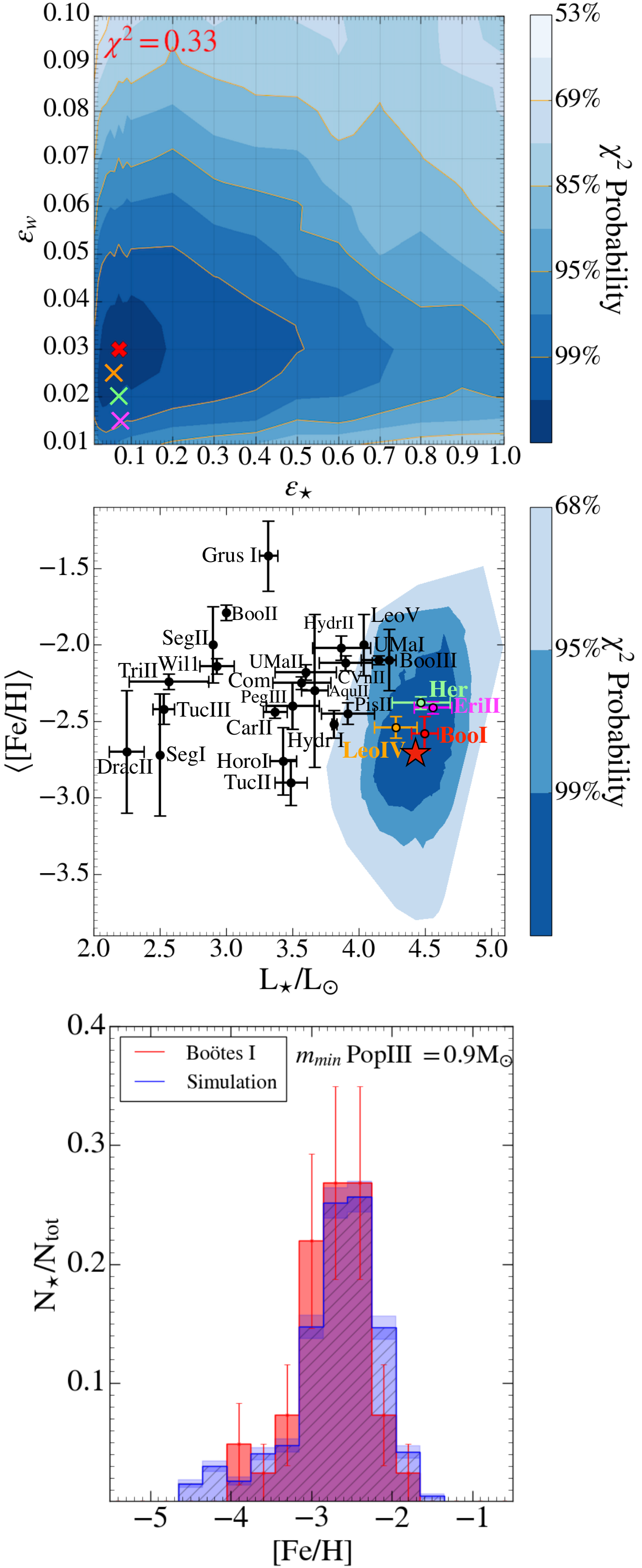}
    \caption{Top panel: $\chi^2$ confidence level contours in the parameter space $(\e, \epsilon_w)$. Red, orange, light green and magenta crosses indicate the minimum of the reduced $\chi^2$ for Bo{\"o}tes I, Leo IV, Hercules and Eridanus II, respectively. Middle panel: Iron-Luminosity relation, colours trace the $\chi^2$ confidence level contours. Bottom panel: comparison between the simulated (blue), and the observed (red) MDF of Bo{\"o}tes I. 
    }
    \label{modelcalcalibration}
\end{figure}

The model includes three free parameters: the infall time, $t_{inf}$, the star formation efficiency, $\epsilon_{\star}$, and the SN wind efficiency, $\epsilon_{w}$. We calibrate our model by comparing with observations, using a statistical approach. In particular, we fix the free parameters of the model to reproduce the following observed properties of Bo{\"o}tes I:
\begin{itemize} 
\item The total luminosity of the galaxy, $\log(\rm L_{\star}/{L_{\odot}}) = 4.5 \pm 0.1$ (\cite{kirby13});
\\
\item The average iron abundance of stars, $\langle \rm [Fe/H] \rangle~= -2.58~\pm~0.43$ (\cite{Lai11}, \cite{norris10});
\\
\item The time interval, counted starting from the first star formation event, needed to form at least the $50\%$ of the total stellar mass  ($Cumulative$ $SFH$= 0.5, $t_{50} \approx (600 \pm 400) \: \rm Myr$, \cite{Brown14});
\\
\item The observed metallicity distribution function (MDF)\footnote{we use 8 points in the observed MDF} (\cite{norris10}, \cite{Lai11}, \cite{gilmore13}).
\end{itemize}
Since PopIII stars have not been observed so far, we calibrated our model by assuming that there are no zero-metallicity stars survivors. Therefore, we set the minimum mass of PopIII stars to $m_{min} = 0.9\msun$.
\\

\noindent By varying $t_{inf}$, $\epsilon_{\star}$, $\epsilon_{w}$ we find the combination that minimises the reduced $\chi^2$. The $\chi^2$ distribution presents a minimum for $(t_{inf}, \epsilon_{\star}, \epsilon_{w}) = (14.5 \: \rm Myr,  0.07, 0.03)$ for which we get $\chi^2 =~0.33$. In the top panel of Fig.\ref{modelcalcalibration} we show the $\chi^{2}$ confidence levels in the parameter space of $\epsilon_{\star}$ and $\epsilon_{w}$ at fixed $t_{inf} = 14.5 \: \rm{Myr}$. We note that  degeneracy exists between the parameters, however the minimum of $\chi^{2}$ is unique.  \\

\noindent  From each combination of $(\epsilon_{\star}, \epsilon_{w} )$, we derive $\langle \rm [Fe/H] \rangle $, and the total luminosity, $ \rm L_{\star}$. The corresponding $\rm Fe-L_{\star}$ relation is shown in the middle panel of Fig.\ref{modelcalcalibration}, where we compare our model results with the observed $\rm \langle[Fe/H] \rangle-L_{\star}$ relation for Bo{\"o}tes~I and other Local Group UFDs (\citet{kirby11}, \citet{Mccon12}, \citet{kirby13}, \citet{Simon19},  \citet{Eridanus}). As we can see, the $\chi^2$ $99\%$ confidence level contour contains three other UFD galaxies: $\rm Hercules$, $\rm Leo \: IV$ and  $ \rm Eridanus \: II$. %
 This means that we expect these galaxies to have experienced an evolution similar to that of Bo{\"o}tes~I and hence to have similar best values of the free parameter ($\epsilon_{\star}, \epsilon_w$). To check this, we re-calibrate the model to fit the observed properties of each one of these three additional UFDs (see Tab.\ref{tab1} for their physical properties). The ($\epsilon_{\star}, \epsilon_w$) values obtained for these UFDs are shown in Fig.\ref{modelcalcalibration} (top panel). We see that in all cases the combinations of $(\epsilon_{\star}, \epsilon_{w})$ are enclosed in the $99\%$ $\chi^2$ confidence level contours of Bo{\"o}tes I.

\noindent Finally, the bottom panel of Fig.\ref{modelcalcalibration} shows the comparison between the observed MDF of Bo{\"o}tes I and the simulated one, which are in good agreement, while the same comparison for Hercules, LeoIV and Eridanus II is in Apppendix \ref{MDFsUFDs}.
\begin{table}
\begin{center}
\begin{tabular}{ |c|c|c|c| } 

\hline
\hline
Galaxy & $\rm \log(L_{\star}/L_{\odot}) $ & $\langle\rm[Fe/H] \rangle$ & $t_{50}(\rm Myr)$   \\
\hline
\hline
Bo{\"o}tes I & $4.5\pm 0.1$ & $-2.58 \pm 0.43 $ & $(600\pm400)$  \\
Hercules & $4.46\pm 0.14$ & $-2.41 \pm 0.04 $ & $(600\pm500)$  \\
Leo IV & $4.28\pm 0.16$ & $-2.54 \pm 0.07 $ & $(400\pm300)$  \\
Eridanus II &  $4.70\pm 0.22$ & $-2.38 \pm 0.04 $ & $(100   \pm \: ...\, )$\\
\hline

\end{tabular}
\caption{\label{tab1} Observed properties of UFDs in the $\chi^2$ $99\%$ confidence level contour of Bo{\"o}tes I. The observed MDFs are shown in Appendix \ref{MDFsUFDs}.}

\end{center}
\end{table}

\section{Modelling the IMF Random Sampling}
\label{samplingdescription}
The data-calibrated model was obtained by assuming to fully populate the IMF (Sec.~\ref{modeldescriprtion}). It is customary to treat the stellar mass distribution, i.e the IMF, like a continuous function where the stellar mass formed is distributed over the whole range of masses.  
When dealing with a large galaxy, or a high star-formation rate in general, it is justified to think of the IMF as a densely sampled probabilistic function. However, when dealing with poorly star-forming systems, the hypothesis of a fully populated IMF breaks down (\cite{kroupa03}, \cite{weidner06}, \cite{Carigi08}, \cite{kroupa11}, \citet{leaman12}, \cite{weidner13}, \cite{debennessuti17}, \cite{applebaum18}). 
It is therefore important to understand how the incomplete sampling of the IMF influences the chemical evolution of poorly star-forming UFDs.

\subsection{Stochastic IMF sampling procedure} \label{sect:stochasticIMF}
\label{imfprocedure}
The Monte Carlo procedure that is used to determine the masses of stars formed in a single burst, $ \rm M_{\star}$, can be described as follows: we choose a normalized IMF (e.g. Larson, Eq.\ref{larson}), and then we build up the probability function, i.e. the cumulative IMF that ranges between 0 and 1. For each randomly extracted number between 0 and 1, we determine the corresponding mass through the probability function. When the extracted random number falls in the mass interval between $ m_{\star}^{i}$ and $m_{\star}^{i+1}$ we assume to form a star with mass $ m_{\star}^{i} = (m_{\star}^{i} + m_{\star}^{i+1})/2$. Random numbers are generated until the total mass of stars equals the stellar mass formed, $\rm M_{\star}=\sum_i m^{i}_{\star} N^{i}_{\star}$, where $\rm N^{i}_{\star}$ is the number of stars with mass $\rm m^{i}_{\star}$. Thus at each time step, we obtain the {\it effective IMF} that can be compared with the theoretical one. Our results of the random sampling are in good agreement with those of \cite{debennessuti17}.
\\ 

  \begin{figure*}
	\includegraphics[width=0.98\textwidth]{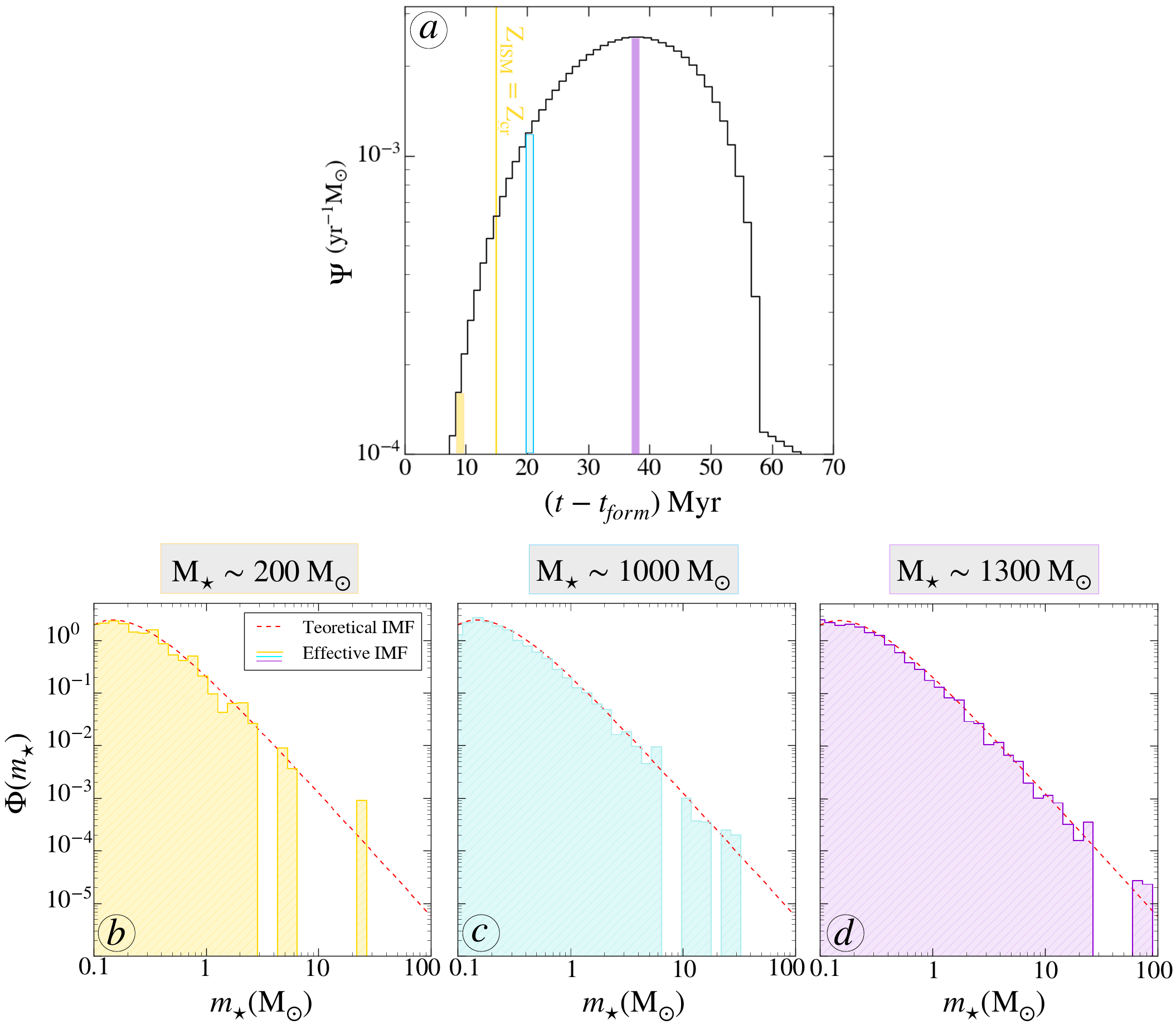}
    \caption{{\it Panel a}: star formation history of Bo{\"o}tes I according to our fiducial model. The vertical yellow line shows the transition between PopIII and PopII stars. The coloured shaded histogram bins underline the star-formation rate in three different evolutionary phases: the early star formation, $(t-t_{form}) \approx 10 \: \rm Myr$, when PopIII star form at a very inefficient rate ($\Psi \approx 2\cdot 10^{-4} \msun \rm yr^{-1}$, yellow);
    and two other stages, $(t-t_{form}) \approx 20-40 \: \rm Myr$, when PopII stars form and the star formation rate is at the average ($\Psi \approx 10^{-3}\msun \rm yr^{-1}$, blue) and the maximum ($\Psi \approx2.5\cdot 10^{-3}\msun \rm yr^{-1}$, purple) value. Panels {\it b)-c)-d)}: for the three evolutionary phases highlighted in {\it panel a)} we compare the normalized  theoretical IMF (dashed line) and the effective IMF (histograms) resulting from our Monte Carlo procedure (Sec.\ref{sect:stochasticIMF}). Note that the IMFs have been normalized to one.
    }
    \label{SFR}
 \end{figure*}

\noindent In Fig.\ref{SFR} we illustrate how this random IMF sampling affects poorly star-forming UFDs when a normal Larson IMF is assumed for both PopIII and PopII/I stars. The top panel shows the predicted SFH of Bo{\"o}tes~I, which illustrates that UFDs have extremely low star-formation rates across their whole evolution ($\Psi < 3 \cdot 10^{-3}\msun \rm yr^{-1}$). The bottom panels show the comparison between the theoretical IMF and the effective one, resulting from the Monte Carlo procedure, for three different evolutionary phases. The lower is~$\Psi$, the worse is the sampling of the theoretical IMF, in particular at the high-mass end. Furthermore, the overall star-formation rate is so low in these small systems, that when the IMF is shifted towards higher masses, $m_{ch}\geq1\msun$, we can only form a few stars around the peak. In those cases, the IMF is poorly populated both at high and low masses (Appendix~\ref{appendixA1}). As we will see in Sec.\ref{impactIMF} this effect is extremely important for the surviving PopIII fraction.
\\

\noindent It should be noted that these results might be affected by the choice of $\rm \Delta t$, since the IMF random sampling strongly depends on the total stellar mass formed in a single time-step. However, we are limited in choosing the time-step resolution: if we want to adequately follow the stellar evolution we need to have $\rm \Delta t < 4 \rm \:  Myr$ (see Sec.\ref{numerical}). On the other hand, the typical time scales for star formation are $\rm \Delta t \geq 0.1$~Myr. We investigated how different choices of $\rm \Delta t$, in the allowed range $[0.1-4]$Myr, affect the chemical evolution of the galaxy. In order to match the global properties of Bo{\"o}tes I for larger (smaller) time-steps we are obliged to choose lower (higher) star formation efficiencies, so the resulting $\Psi$ and fraction of PopIII stars are exactly the same. In other words, the system is ``self-regulated''  by feedback processes, and the results of the IMF random sampling are robust.\\
 
\noindent In conclusion, our results show that even at the peak of the star formation in Bo{\"o}tes I (Fig.\ref{SFR}) there is not a complete convergence between the effective IMF and the theoretical one. In general, we find that the overall mass range can be fully sampled only when $\Psi \gtrsim $ $10^{-1} \: \msun \rm yr^{-1}$, which is never the case in UFDs. Thus, our findings demonstrate that to study UFDs it is fundamental to model the incomplete IMF sampling, which not only affects PopIII star formation but also the formation of ``normal''  PopII/I stars.  
 
\subsection{Impact of the IMF random sampling} \label{sect:imf inpact}
Accounting for the random sampling of the IMF produces two main differences with respect to a fully populated stellar mass distribution. Firstly, the statistical sampling of the IMF produces a more realistic, discrete number of stars for each mass bin. 
Secondly, when the star formation rate is small ($ \lesssim 10^{-3} \: \msun \rm yr^{-1}$), the effective IMF is only populated with a few stars beyond $8\msun$ (see Fig.\ref{SFR}). Since stars with $8\msun \leq$ $m_{\star} \leq 40\msun$ are those that explode as SNe and are responsible for outflows and enrich the ISM with metals, the chemical enrichment history of the UFDs with and without IMF random sampling could be very different.\\
 
\noindent Due to the stochastic nature of the IMF sampling, every time that a star formation event occurs the effective stellar mass distributions can be differently populated, especially at the higher masses. For this reason we exploit a statistical approach and derive the main properties of Bo{\"o}tes I (see Sec.\ref{modelcal}) by averaging among the results of 50 runs of the code and quantifying the scatter among them.

\begin{figure}
	\includegraphics[width=\columnwidth]{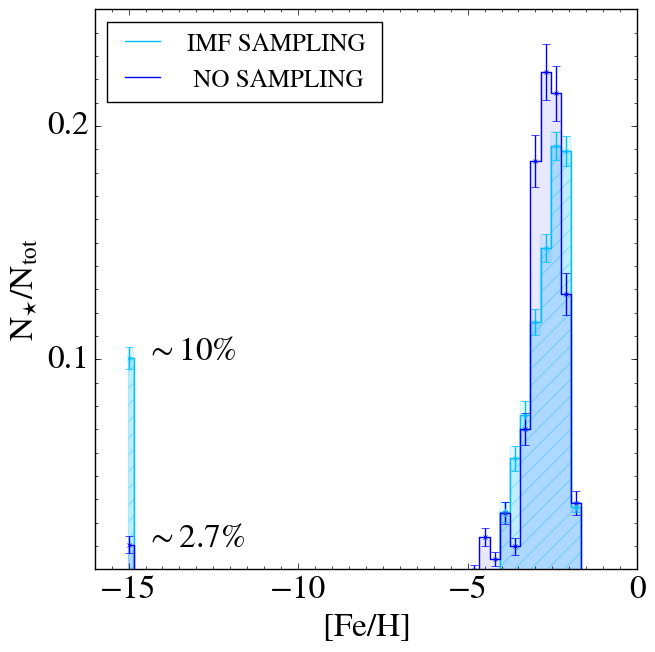}
    \caption{
    Comparison between the MDFs obtained with (light blue) and without (blue, same as Fig.\ref{modelcalcalibration}) IMF random sampling for which we respectively get:
    $ \log(\rm L_{\star}/ L_{\odot}) = 4.1$, $\langle \rm [Fe/H] \rangle = -2.6$, and  $ \log(\rm L_{\star}/ L_{\odot}) = 4.4$, $\langle \rm [Fe/H] \rangle = -2.7$. 
    The MDF obtained by stochastically sampling the IMF has been derived by averaging over 50 runs. The error bars represent the $\pm \sigma$ dispersion among different runs. For the case without the random sampling we show the Poissonian errors.
    \label{MDFconfronto}}
\end{figure}

\label{comparisonandfiducial}
\noindent In Fig.\ref{MDFconfronto} we compare the MDF obtained by assuming a fully populated IMF to that achieved with stochastic IMF sampling. In both cases we assume the same free parameters (Sec. \ref{modelcal}). 
The most important difference between the two MDFs is the number of surviving zero-metallicity stars, which is more than three times higher in the random IMF sampling case, i.e. $\approx 10\%$ of the total number of stars instead of $2.7\%$. 
As we can see in  Fig.\ref{MDFconfronto}, the resulting MDFs are also different in the position and amplitude of the peak. In fact, the MDF obtained with the random IMF sampling has a less pronounced peak that is shifted towards higher iron abundance, $\rm [Fe/H] \approx -2$.
As a consequence the observed properties ($\rm L_{\star}$, and average $[\rm Fe/H]$) are not well reproduced in the random IMF sampling model due to the lower star formation rate (for more details see Appendix~\ref{appendixB}).\\

\noindent We can therefore conclude that when we account for the IMF random sampling in the modelling of poorly star-forming UFD galaxies, their evolution and global properties change.  Hence, the model needs to be  re-calibrated, using the procedure described in Sec.\ref{modelcal}. The new free parameters that minimise the reduced $\chi^2$ are $(t_{inf}, \epsilon_{\star}, \epsilon_{w}) = (14.5 \: \rm Myr,  0.9, 0.03)$.
 Note that $\epsilon_{\star}$ is much larger in this case\footnote{The fraction of gas that at each time-step is converted in stars is given by the ratio between star-formation efficiency and the free-fall time, i.e. $\epsilon_{\star} \cdot dt$/$t_{ff}(z)$ (see Sec.\ref{stelle})}. Because of the poor IMF sampling at high masses, the effect of mechanical feedback driven by SNe is not continuous but intermittent. The impact of SNe explosions is therefore stronger since it is produced by a finite number of SNe instead of fractional numbers (Appendix \ref{appendixB}) indeed,
following the chemical evolution of the galaxy it emerged that the explosion of one, at most two SNe, is enough for the ISM metallicity to exceed $\rm Z^{ISM} = 10^{-3.5} \rm Z_{\odot}$. As a consequence, the star formation rate is more easily damped. Thus to reproduce the the total luminosity of Bo{\"o}tes~I, a higher star-formation efficiency, $\epsilon_{\star}$, is required. With the re-calibrated model, using these new parameters, we get that the fraction of zero-metallicity stars is $\sim 12\%$ with respect to the total survivor stars.

\section{Constraining the PopIII minimum mass}
\label{impactIMF}

\begin{figure}
\includegraphics[width=\columnwidth]{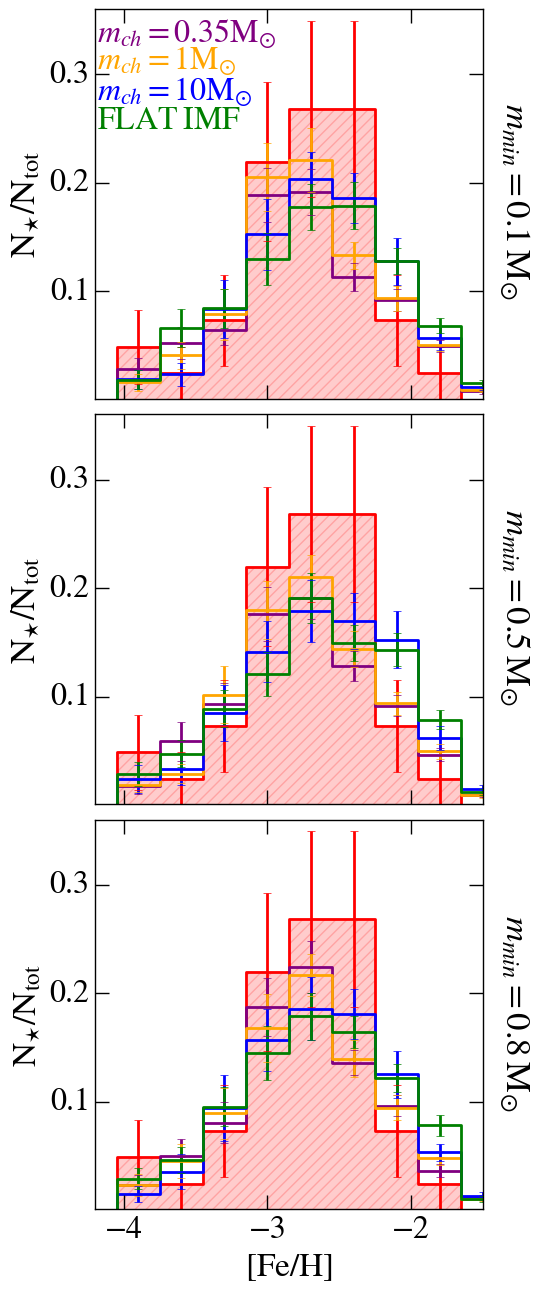}
\caption{The simulated MDFs, obtained by varying the PopIII IMF, in comparison with that of Bo{\"o}tes I (red histogram). In the column we show MDFs obtained by varying the minimum mass, $m_{min}$, of the PopIII stars, while the different colours identify different IMF shapes, i.e. the different characteristic mass, $m_{ch}$.}
\label{MDFs100}
\end{figure} 

So far we have assumed that PopIII stars form according to the present-day stellar  IMF (see Eq. \ref{larson}). With this assumption we estimated that the expected fraction of long-lived PopIII stars in Bo{\"o}tes I is $\sim 12\%$ of the total. We explore how this result changes by varying the PopIII IMF shape, minimum, and maximum mass. As illustrated in Fig.\ref{imfs} we here consider three Larson IMF with different characteristic masses, $m_{ch}$, along with a Flat IMF case (see also Sec.\ref{stellarinitialmassfunction}). 

\subsection{Impact of the PopIII IMF} \label{sect:impact}
Our results show that the observed global properties of Bo{\"o}tes~I are almost entirely determined by PopII/I stars, and that the PopIII IMF has negligible effect on the average $[\rm Fe/H]$ and $\rm L_{\star}$ of the galaxy. PopIII stars form during the first evolutionary phase of Bo{\"o}tes~I, $(t-t_{form}) \sim15$Myr, when the star formation rate is low, $\Psi \lesssim 10^{-3.5} \rm yr^{-1} \msun$ (see Fig.\ref{SFR}). Their contribution to the total stellar mass formed in Bo{\"o}tes I, and therefore to the total $\rm L_{\star}$, is only $\approx 10\%$. Consequently, very few PopIII SNe are formed, both due to their low star formation rate, and the incomplete sampling of the IMF. The overall ISM metal enrichment is therefore dominated by PopII stars.
\noindent Similarly, changing the PopIII IMF only slightly affects the MDF of Bo{\"o}tes I. This is illustrated in Fig.\ref{MDFs100}, where we show the MDFs obtained by varying both the shape and lower mass limit of the PopIII IMF. 
 All the simulated MDFs are in good agreement with that of Bo{\"o}tes I and consistent among each other (i.e. within the dispersion of different runs). This is because the shape of the MDF between $-4 < \rm{[Fe/H] < 1}$ is driven by the PopII/I stars whose IMF has not been changed. \\

\begin{figure}
\includegraphics[width=\columnwidth]{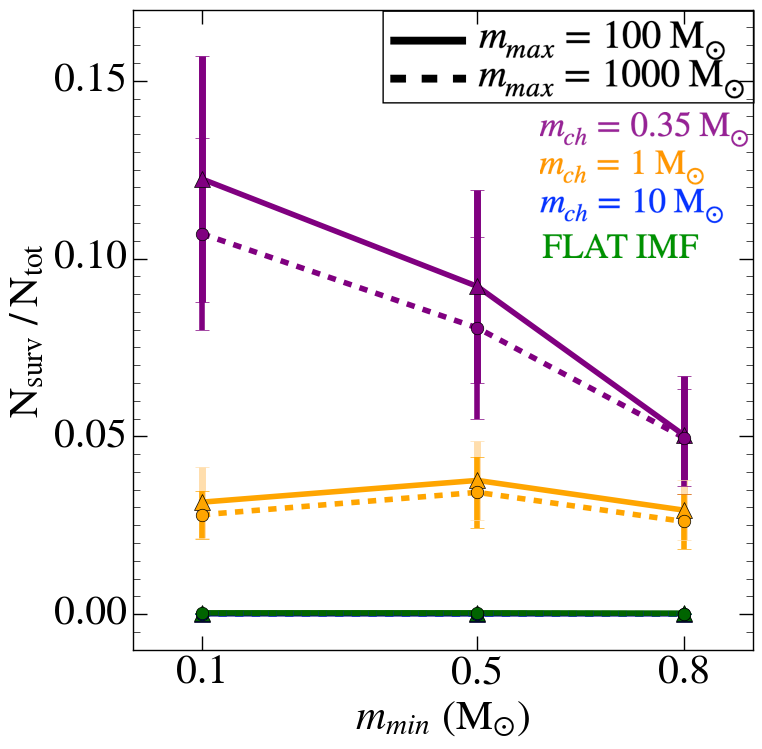}
\caption{The number of PopIII survivors, $\rm N_{surv}$ ($\rm Z_{\star}\le Z_{cr}$), with respect to the total, $\rm N_{tot}$, as a function of the minimum mass of PopIII stars, $m_{min}$, assuming a maximum mass of $m_{max}=100, 1000 \msun$ (solid, dashed lines). The colours represent the different choices of the PopIII IMFs (see Fig.\ref{imfs}).}
\label{popIIIsurv}
\end{figure}

\noindent Conversely, the expected number of PopIII stars changes considerably by varying the PopIII IMF. This is illustrated in Fig.\ref{popIIIsurv}, where we show the number of surviving PopIII stars, $\rm N_{surv}$, with respect to the total number of stars in Bo{\"o}tes I, $ \rm N_{tot}$, for different $m_{min}$, $m_{max}$, and shape of the PopIII IMF.
At fixed $m_{min}$ we can see that $\rm N_{surv}/N_{tot}$ decreases as $m_{ch}$ increases. This is expected since when $m_{ch}$ increases it becomes less likely to form stars with $m _{\star} \le 0.8 \: \msun$, especially when the star formation rate is low ($ \Psi \lesssim 10^{-3} \: \msun \: \rm yr^{-1} $). For this reason, both the Flat PopIII IMF and a Larson PopIII IMF with $m_{ch}=10\: \msun$ give very low $\rm N_{surv}/N_{tot}<0.0001$.\\

\newpage
\noindent By varying $m_{min}$ at fixed $m_{ch}$, we can identify two different trends: for $m_{ch}= 0.35\msun$, $\rm N_{surv}/N_{tot}$ decreases with increasing $m_{min}$, while for all the others IMFs the increase of $m_{min}$ has essentially no effect on the fraction of long-lived PopIII stars.  These trends can be explained as a combination of two effects: the shape of the theoretical IMF, i.e. the higher is the characteristic mass, the more unlikely it is to form low-mass stars; and the incomplete IMF sampling. 
For $m_{ch}=0.35\msun$ the low-mass end of the PopIII IMF, i.e. where the maximum resides, is well populated even when the star formation rate is very low (see Fig.\ref{SFR}). So when we increase $m_{min}$ we are automatically decreasing the number of low-mass PopIII stars that can form and survive until the present day. Increasing $m_{min}$ therefore leads to a decrease in the percentage of survivor PopIII stars. 
Instead for $m_{ch}>1\msun$ and a Flat IMF, when $\Psi \lesssim 10^{-3} \: \msun \rm yr^{-1}$ it is unlikely to form stars with $m_{\star} \le 0.8 \msun$. Hence, by increasing $m_{min}$ the probability to form low-mass stars does not change significally and therefore the percentage of long-lived PopIII stars is roughly constant. \\

\noindent In Fig.\ref{popIIIsurv} we also explore how the fraction of PopIII stars changes if we extend the maximum mass of PopIII stars, $ m_{max}$, up to $1000\msun$. As we can see, $\rm N_{surv} / \: N_{tot}$ is roughly independent of $ m_{max}$. In fact, as we have previously discussed, PopIII stars formed in the first evolutionary phases of the galaxy, when $\Psi \lesssim 10^{-3} \: \msun \rm yr^{-1}$. It follows that the PopIII IMF is badly populated, especially for $m_{\star} \ge 10 \msun$, whatever IMF is chosen (Fig.\ref{SFR} panel a). Hence the probability to form low-mass stars does not change. In conclusion, extending $m_{max}$ up to $1000\msun$ has very little effect on the fraction of PopIII stars that we expect to find in Bo{\"o}tes~I.\\

\noindent Finally, we made the same analysis for the other UFDs that have global properties consistent with our model (Fig.\ref{modelcalcalibration}). The fractions of PopIII star survivors that we expect to find in Hercules, Leo~IV and Eridanus~II are all consistent with those of Bo{\"o}tes~I, within the error bars.

\subsection{Constraining the low-mass end of PopIII stars}
\label{constraining}
We can now exploit the non-detections of metal-free stars in Bo{\"o}tes~I to put constraints on the minimum mass of PopIII stars.
The key question is: how many stars do we need to observe to have a probability\footnote{Assuming that all stars have the same probability of being observed, the probability of not having detected any PopIII survivor stars in the observed sample is: 
$p_{0} =  \frac{(\rm N_{\rm tot}-N_{\rm surv})!(\rm N_{\rm tot} - N_{\rm o})!}{\rm N_{\rm tot}! (\rm N_{\rm tot}-N_{\rm surv} -N_{\rm o})!}$, where $\rm N_{\rm tot}$ is the total number of stars in the galaxy at $z=0$, $\rm N_{surv}$ the expected number of survivor PopIII stars, and $\rm N_{\rm o}$ is the number of observed stars. $1- p_{0}$ is therefore the probability to observe PopIII stars.} of $68\%$, $95\%$ and $99\%$ to catch PopIII stars for a given IMF? \\

\begin{figure}
\includegraphics[width=\columnwidth]{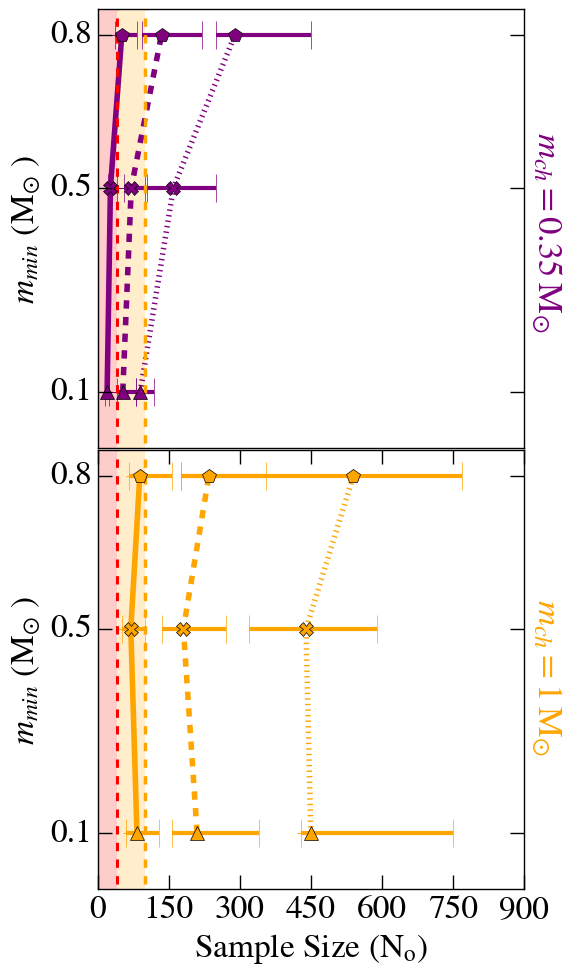}
\caption{The sample size, $\rm N_o$, needed to constrain the PopIII IMF as a function of $m_{min}$ of the PopIII IMF. The different colours represent the PopIII characteristic mass, purple $m_{ch}=0.35$ $\msun$, and orange $m_{ch}=1$ $\msun$. The lines mark the different confident levels: $68\%$ (solid), $95\%$ (dashed), and $99\%$ (dotted).}
\label{limits}
\end{figure}

\noindent Following the statistical approach of \cite{Hartwig15} we  estimate the minimum stellar sample $(\rm N_o)$  needed to be observed in Bo{\"o}tes I, as a function of $m_{min}$ for different PopIII IMFs  to constrain $m_{min}$ (Fig.\ref{limits}). Since no PopIII stars have been discovered so far, all models that predict a $ \rm N_o$ smaller than the total number of stars observed in Bo{\"o}tes I, can be used to constrain $m_{min}$.\\

\noindent 
For $m_{ch}=0.35 \msun$, $\rm N_o$ increases with $m_{min}$ (Fig.\ref{limits}). As  discussed in the previous Sec.\ref{sect:impact} (Fig.\ref{popIIIsurv}), the fraction of PopIII survivors, $\rm N_{surv}$, decreases with increasing $m_{min}$ and so a larger stellar sample is needed to exclude their existence. Instead, $\rm N_o$ is roughly constant for $m_{ch}=1 \msun$ (Fig.\ref{limits}), and this reflects the roughly constant trend of $\rm N_{surv}$ with $m_{min}$ (Fig.\ref{popIIIsurv}).  In Fig.\ref{limits} we only show our findings for $m_{ch}\le1\msun$ because in the case $m_{ch}=10\msun$ and the Flat IMF the sample size required to put constraints is $\sim 10^{4}$, independently of $m_{min}$ and $m_{max}$.\\

\noindent We can now try to get some observationally driven limits on $m_{min}$ for different PopIII IMFs. So far there are only 41 stars in Bo{\"o}tes~I with measured iron abundances. By using these stars  (red area in Fig.\ref{limits}) we can already exclude that PopIII stars formed according to a present-day stellar Larson-type IMF at $68\%$ confidence level. In other words, either $m_{ch} > 0.35\msun$ or $m_{min} > 0.8 \msun$.
We recall that our model is also able to reproduce the properties of the UFD galaxies Hercules, Leo~IV, and Eridanus~II (Fig.\ref{modelcalcalibration}). If we also consider stars that have been observed in these UFD galaxies, the actual sample size goes up to  $\rm N_o =96$ (orange area in Fig.\ref{limits}). Using these additional data we get even tighter constraints, and the present-day stellar IMF is excluded at $95\%$ of confidence level. Furthermore we can assert that $m_{min} > 0.8 \msun$ or $m_{ch}>1 \msun$ at $68\%$ of confidence level.

\section{stronger PopIII IMF constraints}
\label{strong}
By using the non-detection of zero-metallcity stars in Bo{\"o}tes~I, we have constrained the shape and minimum mass of the PopIII IMF (Sec.\ref{impactIMF}). 
However, even if PopIII stars with $m_{\star} \le 0.8 \: \msun$ were able to form and survive until today it does not necessarily imply that they are readily identifiable. Currently, we can obtain the spectra necessary for iron abundance determination, only for the most luminous red giant branch stars in UFD galaxies. Therefore, to quantify and identify PopIII stars that are on the red giant branch, color-magnitude diagrams (CMD) are required. This will allow us to determine how many PopIII stars are bright enough to be observed with current telescopes, but also to  make testable predictions for the new-generation telescopes and instruments (e.g. ELT with the MOSAIC spectrograph (\citet{hammer14}, \citet{evans15})).\\

\begin{figure*}
\includegraphics[width=0.8\textwidth]{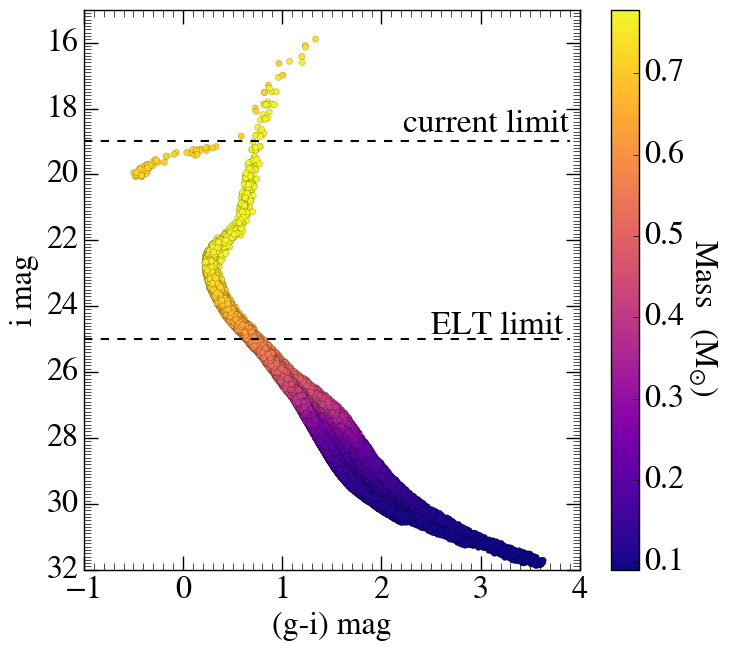}
\caption{The Bo{\"o}tes I simulated CMD, derived using PARSEC CMD generator with SDSS photometric bands. The color bar represents  the mass of the stars in the CMD, while the two dashed horizontal lines indicate respectively the limit magnitude of the current observations of Bo{\"o}tes~I, $\rm i = 19$, and the limit given for MOSAIC on ESO ELT , $\rm i = 25$.}
\label{CMD}
\end{figure*}

\subsection{Simulating the Bo{\"o}tes I CMD}
\label{simCMD}
To simulate the CMD of Bo{\"o}tes~I we used the PARSEC isochrones (\cite{parsec}) and CMD generator\footnote{ available at http://stev.oapd.inaf.it/cmd} adopting the SDSS ugriz photometric system. For the evolutionary tracks we use PARSEC version 1.2S plus COLIBRI $\rm S 35$ (\cite{pastorelli2019}) that add the TP-AGB evolution. For dust we assumed the scaling relations from \cite{marigo08}, the extinction curve of \cite{cardelli89} plus \cite{odonnel94}, and we applied extinction coefficients computed star-by-star. Finally, we selected the Kroupa IMF that, especially in the low-mass end, is the most similar to the Larson IMF assumed in our model (see Sec.~\ref{stellarinitialmassfunction}). Note that PARSEC is limited in metallicity, therefore we assign to metal-free stars the minimum available metallicity of $\rm Z_{\star}= 10^{-7} \: \rm Z_{\odot}$. The synthetic CMD is shown in Fig.\ref{CMD}, where i magnitude has been corrected, taking into account the distance of Bo{\"o}tes~I (\cite{Mccon12}). 
For each star in the CMD we assign a random error in magnitude and color \footnote{We adopt the errors estimated in:\\ http://classic.sdss.org/dr4/algorithms/sdssUBVRITransform.html} in order to simulate real observations. 

\subsection{Can we really catch zero-metallicity stars?}
\label{canwecatch}
The synthetic CMD in Fig.\ref{CMD} shows a clear general trend of stellar mass and luminosity, i.e. the smaller is $m_{\star}$, the fainter is the star. 
The spectroscopic data, currently available for  Bo{\"o}tes I, reach a magnitude $\rm i \sim 19$, for which 41 stars have measured [Fe/H]. At the moment we are thus only able to measure iron and other chemical elements in the more massive, and therefore more luminous stars, $m_{\star} \ge 0.7\msun$. However, to catch potential zero-metallicity stars with $m_{\star} = 0.1 \msun$, observations should reach extremely deep magnitude, i = 32. Indeed, as show in Fig.\ref{CMD}, the planned MOSAIC instrument on the ESO ELT, will be able to reach  $\rm V \approx 25$ at $\rm R \approx15\,000-20\,000$ (\cite{evans15}). This means that most of these very low-mass PopIII stars, if they ever existed, will be invisible even for future generation telescopes.

\begin{figure}
\includegraphics[width=\columnwidth]{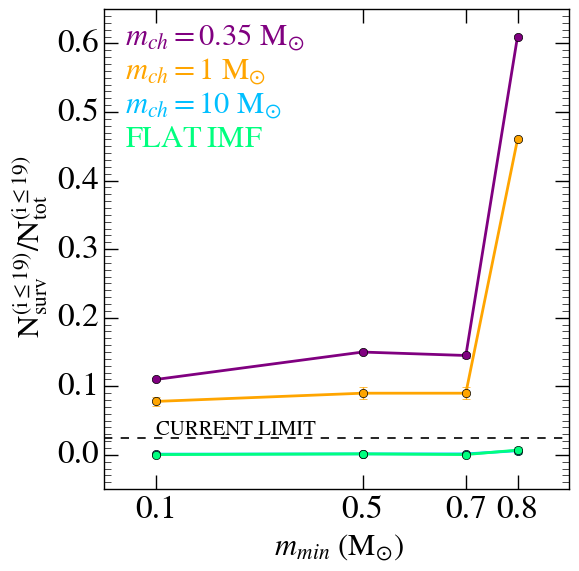}
\caption{The fraction of zero-metallicity stars with respect the truly observable sample of stars as a function of the PopIII minimum mass. Colours identify the different PopIII IMFs, where $m_{max}=100\,\msun$ in all cases.
}
\label{newfraction}
\end{figure}  

\subsection{How to get tighter PopIII IMF constraints}
\label{tigherconstraints}
\noindent In Sec.\ref{constraining} we estimated the fraction of PopIII stars that survive until $z=0$ with respect to the total number of surviving stars in Bo{\"o}tes~I, i.e. all the stars present in the CMD (Fig.\ref{CMD}). What happens if we compute the number of PopIII star survivors within the sample of stars that are realistically observable?\\

\noindent In Fig.\ref{newfraction}, we show the fraction of PopIII survivors in Bo{\"o}tes I, as a function of $m_{min}$, for the current magnitude limit, $\rm i \le 19$, i.e.  $\rm  N^{(\rm i \le 19)}_{\rm surv} / N^{(\rm i \le 19)}_{tot}$. For the two cases, $m_{ch}= 0.35$ and $1\msun$, the new fractions of zero-metallicity stars increase with $m_{min}$, i.e. the trend is inverted with respect to what was found in Fig.\ref{popIIIsurv}. Furthermore at $m_{min} = 0.8 \msun$, the fractions for $m_{ch}= 0.35\msun$ ($m_{ch} = 1\msun$) are one order of magnitude higher than in Fig.\ref{popIIIsurv}, reaching $60\%$ ($45\%$) of the total, instead of $5\%$ ($3\%$). \\

\noindent The reason for these significant differences has to be found in the stellar mass - magnitude relationship, shown in Fig.\ref{CMD}.  In the case of $m_{min} = 0.1\msun$ and $m_{ch} = 0.35\msun$, long-lived PopIII stars are distributed over all magnitudes according to their masses, in the same way as PopII/I stars. The fraction of PopIII survivors at i < 19 is thus comparable to the total fraction (e.g. $12\%$ for  $m_{ch}= 0.35\msun$ in both Fig.\ref{popIIIsurv} and Fig.\ref{newfraction}). However, as $m_{min}$ increases, a higher number of PopIII star survivors is more massive and thus more luminous. This results in a higher fraction of long-lived PopIII stars in the observable region, i.e. $\rm i\le 19$ (corresponding to $m_\star\gtrsim0.7\msun$). Going to deeper magnitudes, in this case, will thus only result in a higher contamination by PopII stars and thus a lower $\rm  N_{\rm surv} / N_{tot}$. 
On the other hand, if $m_{ch}=10\msun$ or if the IMF is flat, the fractions of PopIII star survivors do not vary significantly with respect to Fig.\ref{popIIIsurv} since the probability to form long-lived $m_{\star}<0.8\msun$ PopIII stars is always very low,  independently on $m_{min}$ (see Fig.\ref{imfs} and Appendix \ref{appendixA1}).\\

\noindent Among the 41 stars spectroscopically observed in Bo{\"o}tes I, no long-lived PopIII stars ($\rm Z_{\star}<Z_{cr}=10^{-4.5}Z_{\odot}$) have been identified. We can thus assert that, according to current observations, the probability to have long-lived PopIII stars in  Bo{\"o}tes I  should be $\rm P_{obs} < 1/ N_{tot}= 1/41 \sim 2\%$. Since $\rm P_{obs}$ represents the upper limit of the probability to detect PopIII stars, all models that predict a probability larger than $\rm P_{obs}$ can be discarded. This current (upper) limit is shown in Fig.\ref{newfraction}, where we can appreciate the power of this simple approach: models with  $m_{\star}\leq 0.8\msun$ or $m_{ch}\leq 1\msun$ can be already excluded, which implies that we can put even tighter constraints on the PopIII IMF.\\

\begin{figure}
\includegraphics[width=0.9\columnwidth]{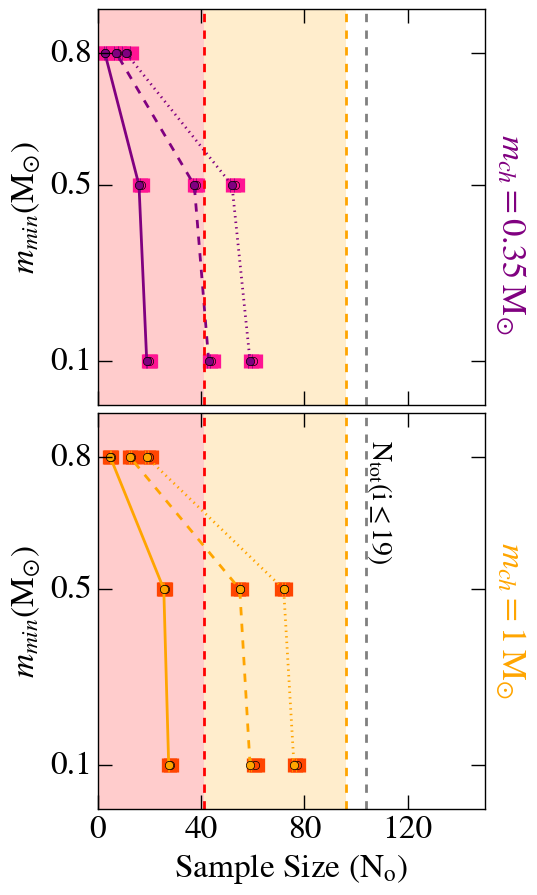}
\caption{The sample sizes, $\rm N_{o}$, needed to constrain the minimum mass, $m_{min}$, of PopIII stars, for $m_{max}=0.35 \msun$ (purple) and  $m_{max}=1 \msun$ (orange), at different confidence levels, for stars with $\rm i \le 19$. The shaded areas display the currently observed stars in Bo{\"o}tes~I (red) and Bo{\"o}tes~I+Hercules+Leo~IV+Eridanus~II (orange). The vertical dashed line in grey delineates the total number of stars with $\rm i\le 19$ in Bo{\"o}tes I. The filled circles represent the results assuming $m_{max}=100\msun$ for PopIII stars, while squares correspond to $m_{max}=1000\msun$.}
\label{NEWCONSTRAINTS}
\end{figure}

 
\noindent In Fig.\ref{NEWCONSTRAINTS} we show the sample size, $\rm N_{o}$, needed to constrain the low-mass end of PopIII IMF at different confidence levels, using the approach from Sec.\ref{constraining}, but now only focusing on stars with $\rm i \leq 19$.
First, we notice that the required stellar sample size is smaller ($\rm N_{o}\leq 100$) with respect to the previous case ($\rm N_{o}\sim 800$, Fig.\ref{limits}). In fact, $\rm N_{o}$ strongly depends on both $\rm N_{surv}^{ (\rm i\leq 19)}$ and $\rm N_{tot}^{\rm (i \leq 19)}$; the higher is the fraction of zero-metallicity stars the lower is $\rm N_{o}$.
By using the current sample of stars observed in Bo{\"o}tes I, we are able to exclude $m_{min}=0.8\msun$ at $99\%$ of confidence level for $m_{ch}=0.35\msun$ and $m_{ch}=1\msun$. Furthermore, by including the data for Hercules, Leo~IV and Eridanus II, we can conclude that PopIII stars should have $m_{min}>0.8 \: \msun $ or $m_{ch} > 1 \msun$ at $99\%$ of confidence level. Note that we are not showing models with $m_{ch} = 10\msun$ and a Flat IMF since $\rm N_{surv}^{(\rm i\leq 19)}/N_{tot}^{\rm (i \leq 19)}\approx 0.2\%$ (see Fig.~\ref{newfraction}) hence even by observing all stars with $\rm i\leq 19$ in Bo{\"o}tes I ($\rm N_{tot}^{( i \leq 19)} \approx 100$) we are unable to constrain $m_{min}$.  

\subsection{How deep should we go?}
\label{howdeep}

\begin{figure}
\includegraphics[width=0.9\columnwidth]{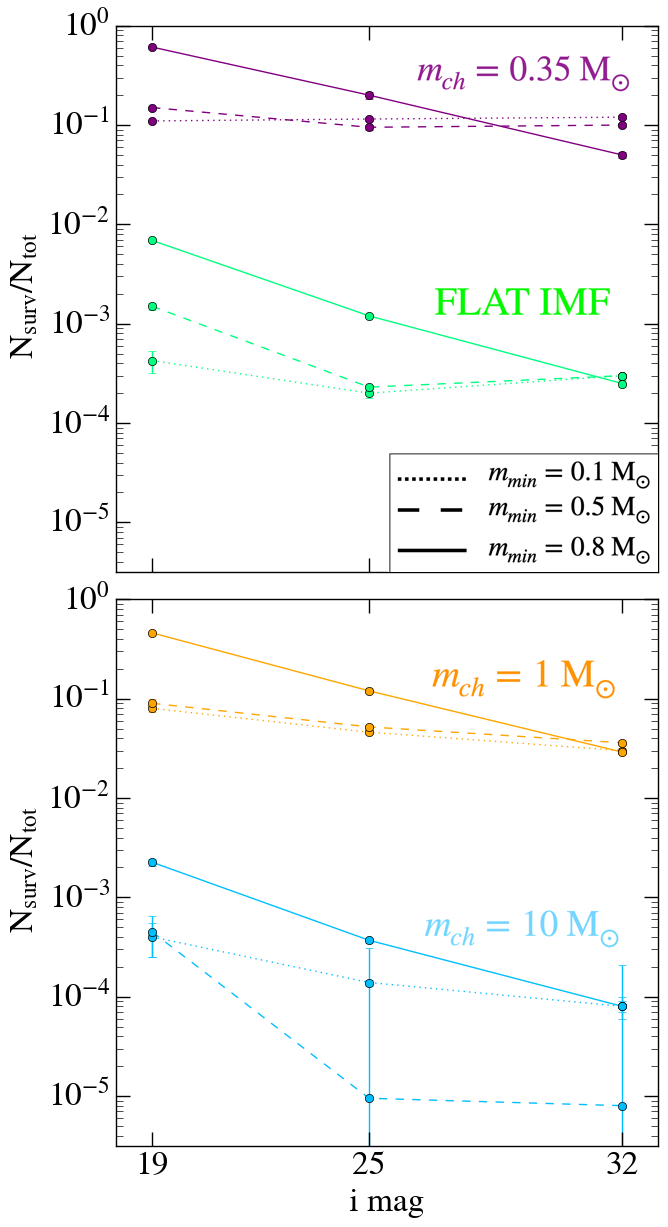}
\caption{Fractions of PopIII survivor stars in  Bo{\"o}tes I as a function of the i magnitude for PopIII IMFs with different shapes and $m_{min}$, assuming $m_{max}=100 \msun$. }
\label{POPIIINEWFRACTIONS}
\end{figure}

\noindent Fig.\ref{POPIIINEWFRACTIONS} shows how the fraction of PopIII stars changes as a function of the observed magnitude. In most cases we see, somewhat counterintuitively, that the fraction of PopIII survivor stars {\it  decreases as we reach fainter magnitudes}. This is because most of the PopIII IMFs tested here are more top-heavy compared to the adopted PopII/I IMF. This means that as we go to fainter magnitudes (and thus reach lower-mass stars), the stellar sample becomes more and more dominated by the normal low-mass PopII/I stars. An exception is when the PopII/I and PopIII IMF is assumed to be the same ($m_{ch}=0.35$ and $m_{min}=0.1$). In that case the trend with limiting magnitude is flat (Fig.\ref{POPIIINEWFRACTIONS}) since both populations are equally distributed in mass (and thus magnitude).\\ 

\noindent In conclusion, a higher limiting magnitude does not imply a higher fraction of PopIII survivors. So why should we go deeper in magnitude? The key point is that for $m_{ch} > 1 \msun$ the predicted fraction of PopIII star survivors is so low ($<0.5\%$) that if we want to constrain $m_{min}$ to at least the $68\%$ of confidence level we need larger stellar samples, which can only be obtained with deeper observations.\\

\begin{figure*}
\includegraphics[width=0.9\textwidth]{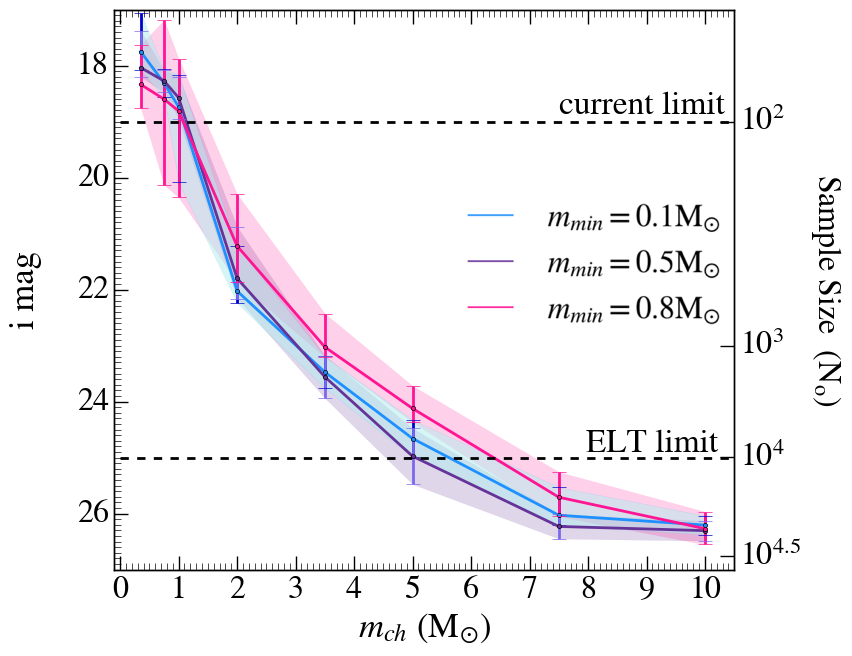}
\caption{The magnitude or equivalently the total number of observed stars in Bo{\"o}tes~I required to constrain the minimum mass of PopIII stars, $m_{min}$, at $68\%$ of confidence level, as a function of $m_{ch}$. Colours delineate models with different $m_{min}$, while the shaded areas represent the $\pm \rm \sigma$ errors.}
\label{nsizemag}
\end{figure*}

\noindent We use the simulated CMD (Fig.\ref{CMD}) to count the number of stars at a given magnitude in Bo{\"o}tes I. The results are illustrated in Fig.\ref{nsizemag}, where we show the limiting magnitude (sample size) required to constrain $m_{min}$ at the $68\%$ confidence level for different $m_{ch}$. Future instruments and telescopes, such as MOSAIC on the ELT, will allow us to measure the metallicity of $10^4$ stars in Bo{\"o}tes I (see Fig.\ref{nsizemag}) and thus either discover a bona fide zero-metallicity star or convert the persisting non-detection into even stronger PopIII IMF constraints: $m_{min}> 0.8 \msun $ or $m_{ch} > 5 \msun$ at $68\%$ confidence level. To get constraints for $m_{ch}=10 \msun$ (or a Flat IMF) we need $\rm N_o \approx10^{4.3}$ stars, which implies to observe $\sim 90\%$ of the stars in Bo{\"o}tes I and thus to reach extremely deep magnitudes, $\rm i=26$. 
 Finally, after simulating the theoretical CMDs for Hercules, Leo~IV and Eridanus~II, thereby accounting for the different distances, we find that with MOSAIC on the ELT we can collect a sample size of $10^{4.2}$ stars at $\rm i\leq 25$, and therefore constrain the PopIII minimum mass, $m_{min } > 0.8 \msun$, for $m_{ch} = 7 \:\msun$ at $68\%$ of confidence level. 
In conclusion, by targeting UFDs with next generation instruments and telescopes we will be able to tightly constrain the minimum mass of PopIII stars, $m_{min}>0.8\msun$, independent of the shape of their IMF.
\subsection{Deeper or wider?}
\label{Deeper or wider}
\begin{figure*}
\includegraphics[width=\textwidth]{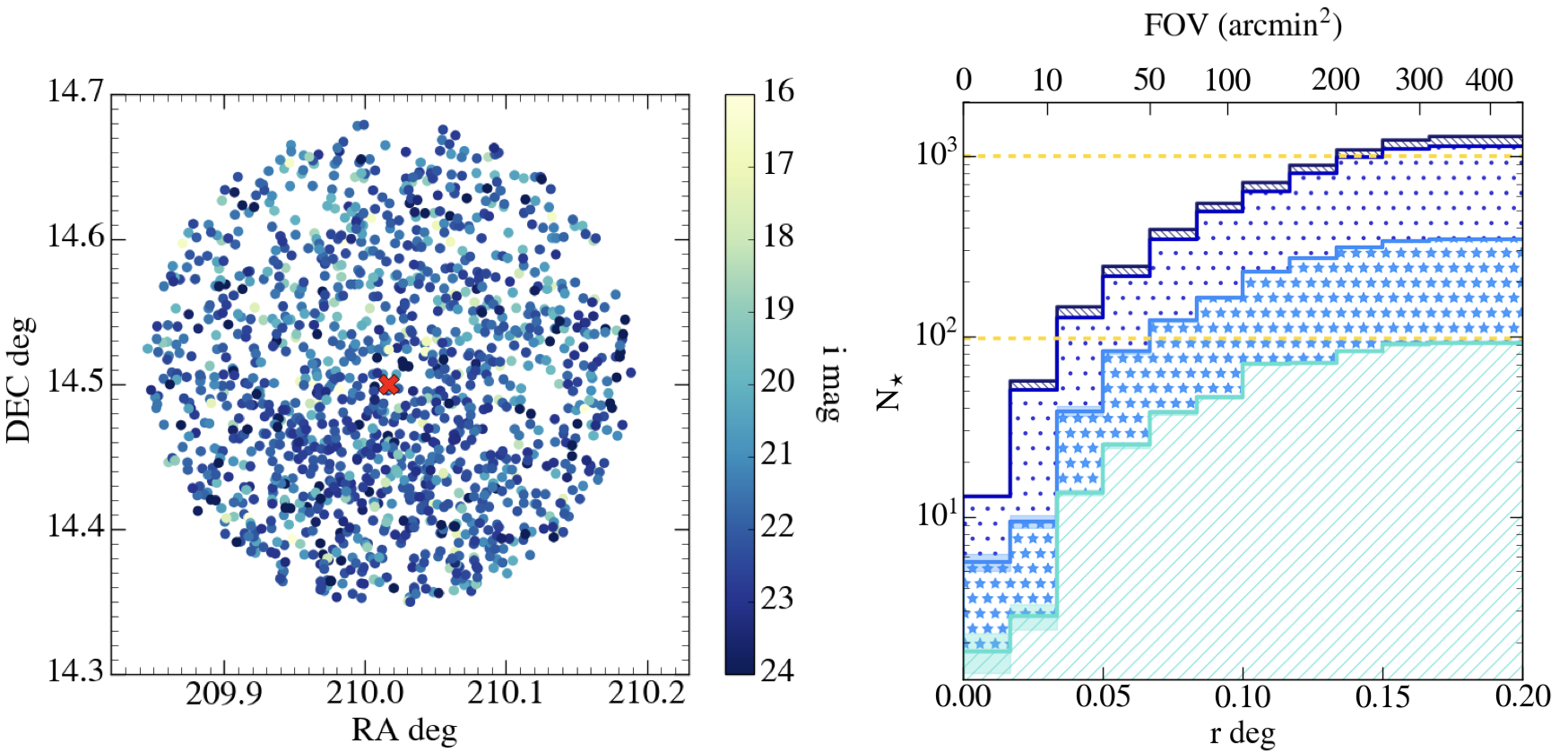}
\caption{Left: the Bo{\"o}tes~I spatial distribution of stars from the simulated CMD. The color bar highlights the different magnitudes, while the red cross identifies the center of mass, $x_{cm}= 210.017$ deg, and $y_{cm}= 14.5$ deg. Right: the number of stars enclosed in a circumference of radius r (bottom) and the corresponding FOV (top), for different limiting magnitudes. These trends were obtained by averaging over 50 realisations of the Bo{\"o}tes I simulated CMD and the corresponding spatial distributions, and the shaded areas represent the $\rm \pm \sigma$ Poisson errors.}
\label{nstars}
\end{figure*}

Combining observations and theoretical predictions for Bo{\"o}tes~I allows us to derive the galaxy's {\it{real}} stellar spatial distribution.
To achieve this goal, we take advantage of the SDSS archive to obtain the observed CMD of Bo{\"o}tes~I.
By comparing the simulated Bo{\"o}tes~I CMD to a real one, we can select member stars from the observed one, and assign RA and Dec to the stars of the simulated CMD. \\
\noindent For each star in the observed CMD we identify the best ``{\it twin}" in the simulated one, i.e. we look for the simulated star with color and magnitude closest to the observed one by minimising the distance in the g-i and i magnitude space.\footnote{$\rm d=\sqrt{ ((g-i)_{obs} - (g-i)_{sim})^2 + (i_{obs} - i_{sim})^2}$} 
We then assign the positions of the observed stars (RA, Dec.) to the simulated ones (Fig.\ref{nstars}, left). Since the errors assigned to the stars in the simulated CMD are random (Sec.\ref{simCMD}), in every run the simulated stars move around their positions in the CMD. Therefore we identify the best twins for each run and then obtain the final results by averaging over 50 runs. The Bo{\"o}tes I SDSS data are limited in magnitude, i = 24, above which the observed CMD is not well populated. Therefore the spatial distribution of the simulated stars, shown in Fig.\ref{nstars}, is limited to $\rm i\le24$. \\


\noindent 
Starting from the center of Bo{\"o}tes~I, we count the number of stars with different limiting magnitudes as a function of the radius and field-of-view (FOV) of the targeted area, see Fig.\ref{nstars} (right). Currently we have spectroscopic data of 41  Bo{\"o}tes~I stars with a limiting magnitude $\rm i \sim 19$. By covering an area $\sim 400 \: \: \rm arcmin^2$, it is possible to observe up to $\approx100$ stars at  $\rm i \leq 19$. However, we can obtain the same sample size by going deeper in magnitude. For example, to reach 100 stars at i = 21 (23), the FOV should be $\approx 50 \:  \rm arcmin^2$ ($\rm 10 \: arcmin^2$). Finally, by reaching limiting magnitude i = 24, and covering the entire FOV of Bo{\"o}tes I, we can obtain a sample of $>10^3$ observed stars, which implies that we can constrain $m_{min}$ for $m_{ch}> 2\msun$ at $68\%$ of confidence level (see Fig.\ref{nsizemag}).

\section{Summary and discussion}
\label{summary}
The aim of this work is to give observationally driven constraints on the Initial Mass Function (IMF) of the first (PopIII) stars and to understand if zero-metallicity low-mass stars ($m<0.8 \msun$) are able to form. We developed, from scratch, a semi-analytical model that describes the evolution of an isolated ultra faint dwarf (UFD) galaxy, from the epoch of its formation until present day. To set the initial conditions of our model we used the results of cosmological simulations (\cite{SS09}, and \cite{SS15}) and we follow their findings by assuming that UFD evolve in isolation, i.e. that they did not experience major merger events. 
The model is data-calibrated, using a statistical approach based on $\chi^2$ analysis. The free parameters of our model have been fixed in order to reproduce the global observed properties of Bo{\"o}tes I, i.e. the total luminosity, $\rm L_{\star}$, the average stellar iron abundance, $\rm [Fe/H]$, the Metallicity Distribution Function (MDF) and the star formation history. \\


\noindent Our key results and the main implications arising from them are:
\begin{itemize}
\item Our analysis of the random sampling of the stellar IMF for $m_{\star}=[0.1-100/1000]\msun$  showed that in poorly star-forming systems like UFD galaxies, $\langle \Psi\rangle \approx 10^{-3}\msun yr^{-1}$, the assumption of a fully populated IMF breaks down (Sec.\ref{sect:stochasticIMF}).\\ 


\item Including the IMF random sampling has a strong impact on the fraction of long-lived PopIII stars in  Bo{\"o}tes I: it increases by a factor of $\sim 3$ with respect to the model in which we assumed to fully populate the IMF (Sec.\ref{sect:imf inpact}).\\

\item We explored how the number of surviving PopIII stars depends on the underlying PopIII IMF, where we tested three Larson-type IMFs ($m_{ch} = [0.35,1,10]\,
\msun$) and a Flat IMF, with different low/high-mass limits (Sec.\ref{sect:impact}). Our results showed that:
    \begin{itemize}
    \item[--] The higher is the characteristic mass, $m_{ch}$, the lower is the number of expected PopIII star survivors.
    \item[--] For $m_{ch} = 0.35 \: \msun$ the fraction of PopIII stars at $ z=0$ strongly depends on the minimum mass, $m_{min}$.
    \item[--] For $m_{ch} > 1 \: \msun$ and for a Flat IMF the number of expected PopIII star survivors is very low (<0.5\%) and roughly independent of $m_{min}$.
    \item[--] The fraction of zero-metallicity stars survivors is not significantly affected by the choice of the maximum mass ($m_{max}=100,1000\,\msun$) of PopIII stars.
    \end{itemize}

\item By taking advantage of the non-detection of zero-metallicity stars among the 41 stars observed in Bo{\"o}tes~I, we can already exclude that PopIII stars form according to a present-day stellar Larson-type IMF, with $m_{ch} = 0.35\msun$ and $m_{min} \le 0.8\msun$, at $68\%$ confidence level. When data for three other UFDs are taken into account (Hercules, Leo~IV and Eridanus II,  also reproduced by our model), we conclude that PopIII stars should have $m_{min} > 0.8\msun$ or $m_{ch} > 1 \msun$ at $68\%$ confidence level (Sec.\ref{constraining}). \\

 
\item Exploiting a synthetic Color Magnitude Diagram (CMD) of the Bo{\"o}tes~I (Sec.\ref{simCMD}), we were able to take into account the mass range of the observed sample. Using Bo{\"o}tes~I data only, this allowed us to conclude that PopIII should not have $m_{min}= 0.8\msun$ at $99\%$ of confidence level  for $m_{ch} =0.35\msun$ and $m_{ch} =1\msun$. If the data of Hercules, Leo~IV and Eridanus II are also taken into account we can conclude that  PopIII stars should have $m_{min} > 0.8\msun$ or $m_{ch} > 1 \: \msun$ at $99\%$ of confidence level (Sec.\ref{canwecatch}-\ref{tigherconstraints}).\\

\item By combining the observed CMD for Bo{\"o}tes~I (based on SDSS data) with the simulated one, we obtained the spatial distribution of stars within the galaxy. 
By reaching $\rm i \leq 24$, and covering the entire Bo{\"o}tes I, we can obtain an observed sample of $>10^3$ stars. As a consequence, if no metal-free star were discovered, we could conclude that PopIII stars should have $m_{min}> 0.8 \msun$ or $m_{ch}> 2\msun$ at $68\%$ of confidence level (Sec.\ref{Deeper or wider}).\\

\item Targeting the four UFDs  Bo{\"o}tes I, Hercules, Leo~IV and Eridanus~II with future generations instruments (such as ELT/MOSAIC) it will be possible to collect a sample of >10\,000 stars. If th non-detection of metal-free stars will be persistent, we will limit $m_{min} > 0.8 \msun$ independent of the PopIII IMF at $68\%$ confidence level (Sec.\ref{howdeep}). \\

\end{itemize}

\noindent In addition, we explored how our results are affected by the choice of critical metallicity value, $\rm Z_{cr}$ that established the transition between PopIII-to-PopII/I stars. The exact value of $\rm Z_{cr}$ is largely uncertain, it is estimated that it can vary between $ \rm Z_{cr}=(10 ^{-4.5 \pm 1}) \: Z_{\odot}$ (\cite{bromm01}, \cite{Schneider2003}, \cite{omukai05}). However we have not found any dependence of the fractions of zero-metallicity stars on critical metallicity. In fact, following the chemical evolution of the galaxy it emerged that the explosion of one, at most two SNe, is enough for the metallicity to exceed $10^{-3.5} \: \rm Z_{\odot}$. Therefore we move from a metal-free environment with $\rm Z_{ISM}= 0 $, to one with $\rm Z_{ISM}  > 10^{-3.5} \: Z_{\odot}$. Hence our results are not affected by the choice of the $\rm Z_{cr}$ value.\\

\noindent There have been several theoretical works (e.g. \cite{vincenzo14}, \cite{romano14}) that model the evolution of  Bo{\"o}tes~I since it is one of the observationally best studied UFD galaxy. However, none of these models take into account the stochastic sampling of the IMF. Furthermore, none of these works have investigated the expected fraction of PopIII star survivors in  Bo{\"o}tes~I. 
In agreement with our findings, \cite{vincenzo14} concluded that the star formation efficiency of  Bo{\"o}tes~I was very small. This is also consistent with studies of Bo{\"o}tes~I in cosmological context (e.g.~\cite{SS09}, \cite{Bovill09Ricotti} \cite{romano14}, \cite{BH2015}, \cite{SS15}). \\

\noindent One of the key points in this work was to estimate the fraction of PopIII star survivors in UFD galaxies. The same was done by \cite{magg17}, who derived the percentage of surviving PopIII that are expected to be found in our Milky Way and its satellites. Unfortunately we cannot make a direct comparison, since they used a logarithmically Flat IMF. However our results for $m_{ch} = 10 \msun $ and the case of Flat IMF predict a PopIII fraction consistent with their findings.\\

\noindent Our methods allow us to estimate the minimum stellar sample size required to put observationally driven constraints on the PopIII IMF. This approach was used by \cite{Hartwig15}, for halo/bulge stars of our Milky Way. They found that the expected number of PopIII star survivors decreases as $m_{min}$ increases assuming a logarithmically Flat PopIII IMF. Our findings have the same trend for IMF with $m_{ch}= 0.35 \msun$. However for the other IMFs ($m_{ch} = 10 \msun$ and Flat IMF) the fraction of PopIII survivors is extremely low ($<0.01\%$) regardless of the minimum mass. 
The reason is twofold: i) the IMF shape, which makes it very unlikely to form low-mass stars (Fig.\ref{imfs}); ii) the random sampling of the IMF which strongly affects the population of the lower mass-end (see Fig.\ref{samplingmch}), i.e. the number of stars with $m_{\star} \le 0.8 \: \msun$. In fact, for $m_{ch}\ge 1\msun$ and for Flat IMF it is unlikely to form long-lived PopIII stars independently of the minimum mass.
Furthermore,  \cite{Hartwig15} found that to conclude that PopIII stars should have $m_{\star} \ge 0.8 \: \msun$ at a $68\%$ confidence level, we need to observe  $\rm N_{o} \approx 10^6$  stars in the Milky Way stellar halo. However, with our method, even tighter constraints can be put on the PopIII IMF by only using $\rm N_{o} \approx 10^2$ stars.\\

\noindent For the first time, we have performed an in-depth analyses of four UFDs, Bo{\"o}tes I, Hercules, Leo~IV and Eridanus~II with the goal to constrain the IMF of PopIII stars. By developing a chemical evolution model of isolated UFDs, and combining observational constraints with a statistical approach we have shown that old and pristine UFDs are the ideal systems to understand the physical properties of the first stars. Our study has revealed that it is crucial to take into account the incomplete sampling of the IMF in these poorly star-forming systems, to realistically represent their stellar populations. This is especially relevant for the earliest star formation, where PopIII stars formed very inefficiently for a limited time. Our novel approach of connecting our models with real data through synthesizing the CMD of Bo{\"o}tes I, has revealed that taking the mass distribution of the observed stars into account, allows us to put even stronger constraints on the $m_{min}$ and the shape of the PopIII IMF. In this work, with only  96 observed stars in Bo{\"o}tes I, Hercules, Leo~IV and Eridanus~II we can derive at a $99\%$ confidence level  that $m_{min} > 0.8 \: \msun$ or $m_{ch} > 1\msun$. Thus, we conclude that PopIII stars were typically more massive than those that form today, and the mass distribution of the first stars was fundamentally different from stars at later times. Finally, we showed that future telescopes will be able to get much larger samples (>$10^4$~stars), which will provide even better understanding of the nature of the very first stars in the universe.

\section*{Acknowledgements}
We thank the anonymous referee for the useful and constructive comments.
We thank Eline Tolstoy and the Cosmology group at the Scuola Normale Superiore of Pisa for insightful comments on the early version of this work. SS thanks Piercarlo Bonifacio for inspiring discussions, which triggered the study of the synthetic CMD of Bo{\"o}tes~I along with the theoretical predictions for future generation telescopes. We acknowledge support from the ERC Starting Grant NEFERTITI H2020/808240. SS acknowledges support from the PRIN-MIUR2017, The quest for the first stars, prot. n. 2017T4ARJ5. 

\section*{Data AVAILABILITY}
The authors confirm that the data supporting the findings of this study are available within the articles: \cite{norris10}, \cite{Lai11}, \cite{gilmore13} for the Bo{\"o}tes~I MDF, \citet{kirby11}, \citet{kirby13} for $\langle\rm[Fe/H]\rangle - \rm L_{\star}$ relation for UFDs.
The data that support the findings of this study are available on request from the corresponding author MR.



\bibliographystyle{mnras}
\bibliography{UFDstheMinimumMassofTheFirstStars} 

\begin{thebibliography}{}
\makeatletter
\relax
\def\mn@urlcharsother{\let\do\@makeother \do\$\do\&\do\#\do\^\do\_\do\%\do\~}
\def\mn@doi{\begingroup\mn@urlcharsother \@ifnextchar [ {\mn@doi@}
  {\mn@doi@[]}}
\def\mn@doi@[#1]#2{\def\@tempa{#1}\ifx\@tempa\@empty \href
  {http://dx.doi.org/#2} {doi:#2}\else \href {http://dx.doi.org/#2} {#1}\fi
  \endgroup}
\def\mn@eprint#1#2{\mn@eprint@#1:#2::\@nil}
\def\mn@eprint@arXiv#1{\href {http://arxiv.org/abs/#1} {{\tt arXiv:#1}}}
\def\mn@eprint@dblp#1{\href {http://dblp.uni-trier.de/rec/bibtex/#1.xml}
  {dblp:#1}}
\def\mn@eprint@#1:#2:#3:#4\@nil{\def\@tempa {#1}\def\@tempb {#2}\def\@tempc
  {#3}\ifx \@tempc \@empty \let \@tempc \@tempb \let \@tempb \@tempa \fi \ifx
  \@tempb \@empty \def\@tempb {arXiv}\fi \@ifundefined
  {mn@eprint@\@tempb}{\@tempb:\@tempc}{\expandafter \expandafter \csname
  mn@eprint@\@tempb\endcsname \expandafter{\@tempc}}}

\bibitem[\protect\citeauthoryear{{Abel}, {Bryan}  \& {Norman}}{{Abel}
  et~al.}{2002}]{abel02}
{Abel} T.,  {Bryan} G.~L.,   {Norman} M.~L.,  2002, \mn@doi [Science]
  {10.1126/science.295.5552.93}, \href
  {https://ui.adsabs.harvard.edu/abs/2002Sci...295...93A} {295, 93}

\bibitem[\protect\citeauthoryear{{Aguado}, {Gonz{\'a}lez Hern{\'a}ndez},
  {Allende Prieto}  \& {Rebolo}}{{Aguado} et~al.}{2019}]{aguado19}
{Aguado} D.~S.,  {Gonz{\'a}lez Hern{\'a}ndez} J.~I.,  {Allende Prieto} C.,
  {Rebolo} R.,  2019, \mn@doi [\apjl] {10.3847/2041-8213/ab1076}, \href
  {https://ui.adsabs.harvard.edu/abs/2019ApJ...874L..21A} {874, L21}

\bibitem[\protect\citeauthoryear{Anders \& Grevesse}{Anders \&
  Grevesse}{1989}]{anders}
Anders E.,  Grevesse N.,  1989, Geochimica et Cosmochimica acta, 53, 197

\bibitem[\protect\citeauthoryear{Aoki, Tominaga, Beers, Honda  \& Lee}{Aoki
  et~al.}{2014}]{aoki14}
Aoki W.,  Tominaga N.,  Beers T.,  Honda S.,   Lee Y.,  2014, Science, 345, 912

\bibitem[\protect\citeauthoryear{Applebaum, Brooks, Quinn  \&
  Christensen}{Applebaum et~al.}{2018}]{applebaum18}
Applebaum E.,  Brooks A.~M.,  Quinn T.~R.,   Christensen C.~R.,  2018, arXiv
  preprint arXiv:1811.00022

\bibitem[\protect\citeauthoryear{Barkana \& Loeb}{Barkana \&
  Loeb}{2001}]{Bark01}
Barkana R.,  Loeb A.,  2001, Physics reports, 349, 125

\bibitem[\protect\citeauthoryear{Beers \& Christlieb}{Beers \&
  Christlieb}{2005}]{beers05}
Beers T.~C.,  Christlieb N.,  2005, Annu. Rev. Astron. Astrophys., 43, 531

\bibitem[\protect\citeauthoryear{{Bland-Hawthorn}, {Sutherland}  \&
  {Webster}}{{Bland-Hawthorn} et~al.}{2015}]{BH2015}
{Bland-Hawthorn} J.,  {Sutherland} R.,   {Webster} D.,  2015, \mn@doi [\apj]
  {10.1088/0004-637X/807/2/154}, \href
  {https://ui.adsabs.harvard.edu/abs/2015ApJ...807..154B} {807, 154}

\bibitem[\protect\citeauthoryear{{Bonifacio} et~al.,}{{Bonifacio}
  et~al.}{2018}]{bonifacio18}
{Bonifacio} P.,  et~al., 2018, \mn@doi [\aap] {10.1051/0004-6361/201732320},
  \href {https://ui.adsabs.harvard.edu/abs/2018A&A...612A..65B} {612, A65}

\bibitem[\protect\citeauthoryear{{Bonifacio} et~al.,}{{Bonifacio}
  et~al.}{2019}]{Bonifacio19}
{Bonifacio} P.,  et~al., 2019, \mn@doi [\mnras] {10.1093/mnras/stz1378}, \href
  {https://ui.adsabs.harvard.edu/abs/2019MNRAS.487.3797B} {487, 3797}

\bibitem[\protect\citeauthoryear{Bovill \& Ricotti}{Bovill \&
  Ricotti}{2009}]{Bovill09Ricotti}
Bovill M.~S.,  Ricotti M.,  2009, The Astrophysical Journal, 693, 1859

\bibitem[\protect\citeauthoryear{{Bressan}, {Marigo}, {Girardi}, {Salasnich},
  {Dal Cero}, {Rubele}  \& {Nanni}}{{Bressan} et~al.}{2012}]{parsec}
{Bressan} A.,  {Marigo} P.,  {Girardi} L.,  {Salasnich} B.,  {Dal Cero} C.,
  {Rubele} S.,   {Nanni} A.,  2012, \mn@doi [\mnras]
  {10.1111/j.1365-2966.2012.21948.x}, \href
  {https://ui.adsabs.harvard.edu/abs/2012MNRAS.427..127B} {427, 127}

\bibitem[\protect\citeauthoryear{Bromm}{Bromm}{2013}]{Bromm13}
Bromm V.,  2013, Reports on Progress in Physics, 76, 112901

\bibitem[\protect\citeauthoryear{Bromm, Ferrara, Coppi  \& Larson}{Bromm
  et~al.}{2001}]{bromm01}
Bromm V.,  Ferrara A.,  Coppi P.,   Larson R.,  2001, Monthly Notices of the
  Royal Astronomical Society, 328, 969

\bibitem[\protect\citeauthoryear{{Bromm}, {Coppi}  \& {Larson}}{{Bromm}
  et~al.}{2002}]{bromm02}
{Bromm} V.,  {Coppi} P.~S.,   {Larson} R.~B.,  2002, \mn@doi [\apj]
  {10.1086/323947}, \href
  {https://ui.adsabs.harvard.edu/abs/2002ApJ...564...23B} {564, 23}

\bibitem[\protect\citeauthoryear{Brown et~al.,}{Brown et~al.}{2014}]{Brown14}
Brown T.~M.,  et~al., 2014, The Astrophysical Journal, 796, 91

\bibitem[\protect\citeauthoryear{Cardelli, Clayton  \& Mathis}{Cardelli
  et~al.}{1989}]{cardelli89}
Cardelli J.~A.,  Clayton G.~C.,   Mathis J.~S.,  1989, The Astrophysical
  Journal, 345, 245

\bibitem[\protect\citeauthoryear{Carigi \& Hernandez}{Carigi \&
  Hernandez}{2008}]{Carigi08}
Carigi L.,  Hernandez X.,  2008, Monthly Notices of the Royal Astronomical
  Society, 390, 582

\bibitem[\protect\citeauthoryear{Chiti et~al.,}{Chiti et~al.}{2021}]{chiti21}
Chiti A.,  et~al., 2021, Nature Astronomy, pp~1--9

\bibitem[\protect\citeauthoryear{Clark, Glover, Smith, Greif, Klessen  \&
  Bromm}{Clark et~al.}{2011}]{clark11}
Clark P.~C.,  Glover S.~C.,  Smith R.~J.,  Greif T.~H.,  Klessen R.~S.,   Bromm
  V.,  2011, Science, 331, 1040

\bibitem[\protect\citeauthoryear{Diemand, Madau  \& Moore}{Diemand
  et~al.}{2005}]{diemand05}
Diemand J.,  Madau P.,   Moore B.,  2005, Monthly Notices of the Royal
  Astronomical Society, 364, 367

\bibitem[\protect\citeauthoryear{{Dopcke}, {Glover}, {Clark}  \&
  {Klessen}}{{Dopcke} et~al.}{2013}]{dopcke13}
{Dopcke} G.,  {Glover} S. C.~O.,  {Clark} P.~C.,   {Klessen} R.~S.,  2013,
  \mn@doi [\apj] {10.1088/0004-637X/766/2/103}, \href
  {https://ui.adsabs.harvard.edu/abs/2013ApJ...766..103D} {766, 103}

\bibitem[\protect\citeauthoryear{Evans et~al.,}{Evans et~al.}{2015}]{evans15}
Evans C.,  et~al., 2015, arXiv preprint arXiv:1501.04726

\bibitem[\protect\citeauthoryear{Frebel}{Frebel}{2010}]{Frebel10}
Frebel A.,  2010, Astronomische Nachrichten, 331, 474

\bibitem[\protect\citeauthoryear{Frebel \& Norris}{Frebel \&
  Norris}{2015}]{Frebel15}
Frebel A.,  Norris J.~E.,  2015, Annual Review of Astronomy and Astrophysics,
  53, 631

\bibitem[\protect\citeauthoryear{Frebel, Norris, Gilmore  \& Wyse}{Frebel
  et~al.}{2016}]{Frebel16}
Frebel A.,  Norris J.~E.,  Gilmore G.,   Wyse R.~F.,  2016, The Astrophysical
  Journal, 826, 110

\bibitem[\protect\citeauthoryear{Freeman \& Bland-Hawthorn}{Freeman \&
  Bland-Hawthorn}{2002}]{freeman02}
Freeman K.,  Bland-Hawthorn J.,  2002, Annual Review of Astronomy and
  Astrophysics, 40, 487

\bibitem[\protect\citeauthoryear{{Gallart} et~al.,}{{Gallart}
  et~al.}{2021}]{Eridanus}
{Gallart} C.,  et~al., 2021, arXiv e-prints, \href
  {https://ui.adsabs.harvard.edu/abs/2021arXiv210104464G} {p. arXiv:2101.04464}

\bibitem[\protect\citeauthoryear{Gilmore, Norris, Monaco, Yong, Wyse  \&
  Geisler}{Gilmore et~al.}{2013}]{gilmore13}
Gilmore G.,  Norris J.~E.,  Monaco L.,  Yong D.,  Wyse R.~F.,   Geisler D.,
  2013, The Astrophysical Journal, 763, 61

\bibitem[\protect\citeauthoryear{Gratton, Sneden  \& Carretta}{Gratton
  et~al.}{2004}]{gratton04}
Gratton R.,  Sneden C.,   Carretta E.,  2004, Annu. Rev. Astron. Astrophys.,
  42, 385

\bibitem[\protect\citeauthoryear{Greif, Springel, White, Glover, Clark, Smith,
  Klessen  \& Bromm}{Greif et~al.}{2011}]{greif11}
Greif T.,  Springel V.,  White S.,  Glover S.,  Clark P.,  Smith R.,  Klessen
  R.,   Bromm V.,  2011, arXiv preprint arXiv:1101.5491

\bibitem[\protect\citeauthoryear{{Hammer} et~al.,}{{Hammer}
  et~al.}{2014}]{hammer14}
{Hammer} F.,  et~al., 2014, in {Ramsay} S.~K.,  {McLean} I.~S.,   {Takami} H.,
  eds,  Society of Photo-Optical Instrumentation Engineers (SPIE) Conference
  Series Vol. 9147, Ground-based and Airborne Instrumentation for Astronomy V.
  p. 914727, \mn@doi{10.1117/12.2055148}

\bibitem[\protect\citeauthoryear{Hartwig, Bromm, Klessen  \& Glover}{Hartwig
  et~al.}{2015}]{Hartwig15}
Hartwig T.,  Bromm V.,  Klessen R.~S.,   Glover S.~C.,  2015, Monthly Notices
  of the Royal Astronomical Society, 447, 3892

\bibitem[\protect\citeauthoryear{Heger \& Woosley}{Heger \&
  Woosley}{2002}]{heger02}
Heger A.,  Woosley S.~E.,  2002, The Astrophysical Journal, 567, 532

\bibitem[\protect\citeauthoryear{{Hirano} \& {Bromm}}{{Hirano} \&
  {Bromm}}{2017}]{HiranoBromm2017}
{Hirano} S.,  {Bromm} V.,  2017, \mn@doi [\mnras] {10.1093/mnras/stx1220},
  \href {https://ui.adsabs.harvard.edu/abs/2017MNRAS.470..898H} {470, 898}

\bibitem[\protect\citeauthoryear{Hirano, Hosokawa, Yoshida, Umeda, Omukai,
  Chiaki  \& Yorke}{Hirano et~al.}{2014}]{Hirano14}
Hirano S.,  Hosokawa T.,  Yoshida N.,  Umeda H.,  Omukai K.,  Chiaki G.,
  Yorke H.~W.,  2014, The Astrophysical Journal, 781, 60

\bibitem[\protect\citeauthoryear{Hosokawa, Omukai, Yoshida  \& Yorke}{Hosokawa
  et~al.}{2011}]{hosokawa11}
Hosokawa T.,  Omukai K.,  Yoshida N.,   Yorke H.~W.,  2011, Science, 334, 1250

\bibitem[\protect\citeauthoryear{{Ishigaki}, {Aoki}, {Arimoto}  \&
  {Okamoto}}{{Ishigaki} et~al.}{2014}]{ishigaki14}
{Ishigaki} M.~N.,  {Aoki} W.,  {Arimoto} N.,   {Okamoto} S.,  2014, \mn@doi
  [\aap] {10.1051/0004-6361/201322796}, \href
  {https://ui.adsabs.harvard.edu/abs/2014A&A...562A.146I} {562, A146}

\bibitem[\protect\citeauthoryear{Ishigaki, Tominaga, Kobayashi  \&
  Nomoto}{Ishigaki et~al.}{2018}]{ishigaki18}
Ishigaki M.~N.,  Tominaga N.,  Kobayashi C.,   Nomoto K.,  2018, The
  Astrophysical Journal, 857, 46

\bibitem[\protect\citeauthoryear{{Ji}, {Frebel}, {Ezzeddine}  \& {Casey}}{{Ji}
  et~al.}{2016}]{ji16}
{Ji} A.~P.,  {Frebel} A.,  {Ezzeddine} R.,   {Casey} A.~R.,  2016, \mn@doi
  [\apjl] {10.3847/2041-8205/832/1/L3}, \href
  {https://ui.adsabs.harvard.edu/abs/2016ApJ...832L...3J} {832, L3}

\bibitem[\protect\citeauthoryear{Kere{\v{s}}, Katz, Weinberg  \&
  Dav{\'e}}{Kere{\v{s}} et~al.}{2005}]{Ker15}
Kere{\v{s}} D.,  Katz N.,  Weinberg D.~H.,   Dav{\'e} 2005, Monthly Notices of
  the Royal Astronomical Society, 363, 2

\bibitem[\protect\citeauthoryear{{Kirby}, {Martin}  \& {Finlator}}{{Kirby}
  et~al.}{2011}]{kirby11}
{Kirby} E.~N.,  {Martin} C.~L.,   {Finlator} K.,  2011, \mn@doi [The
  Astrophysical Journal] {10.1088/2041-8205/742/2/L25}, \href
  {https://ui.adsabs.harvard.edu/abs/2011ApJ...742L..25K} {742, L25}

\bibitem[\protect\citeauthoryear{Kirby, Cohen, Guhathakurta, Cheng, Bullock  \&
  Gallazzi}{Kirby et~al.}{2013}]{kirby13}
Kirby E.~N.,  Cohen J.~G.,  Guhathakurta P.,  Cheng L.,  Bullock J.~S.,
  Gallazzi A.,  2013, The Astrophysical Journal, 779, 102

\bibitem[\protect\citeauthoryear{Klessen, Krumholz  \& Heitsch}{Klessen
  et~al.}{2011}]{klessen11}
Klessen R.~S.,  Krumholz M.~R.,   Heitsch F.,  2011, Advanced Science Letters,
  4, 258

\bibitem[\protect\citeauthoryear{Kroupa \& Weidner}{Kroupa \&
  Weidner}{2003}]{kroupa03}
Kroupa P.,  Weidner C.,  2003, The Astrophysical Journal, 598, 1076

\bibitem[\protect\citeauthoryear{Kroupa, Weidner, Pflamm-Altenburg, Thies,
  Dabringhausen, Marks  \& Maschberger}{Kroupa et~al.}{2011}]{kroupa11}
Kroupa P.,  Weidner C.,  Pflamm-Altenburg J.,  Thies I.,  Dabringhausen J.,
  Marks M.,   Maschberger T.,  2011, arXiv preprint arXiv:1112.3340

\bibitem[\protect\citeauthoryear{Lai, Lee, Bolte, Lucatello, Beers, Johnson,
  Sivarani  \& Rockosi}{Lai et~al.}{2011}]{Lai11}
Lai D.~K.,  Lee Y.~S.,  Bolte M.,  Lucatello S.,  Beers T.~C.,  Johnson J.~A.,
  Sivarani T.,   Rockosi C.~M.,  2011, The Astrophysical Journal, 738, 51

\bibitem[\protect\citeauthoryear{Larson}{Larson}{1998}]{Lars98}
Larson P.~L.,  1998, Gaia, 15, 389

\bibitem[\protect\citeauthoryear{Leaman}{Leaman}{2012}]{leaman12}
Leaman R.,  2012, The Astronomical Journal, 144, 183

\bibitem[\protect\citeauthoryear{Li, Tan  \& Zhao}{Li et~al.}{2018}]{li18}
Li H.,  Tan K.,   Zhao G.,  2018, The Astrophysical Journal Supplement Series,
  238, 16

\bibitem[\protect\citeauthoryear{{Machida}, {Matsumoto}  \&
  {Inutsuka}}{{Machida} et~al.}{2008}]{machida08}
{Machida} M.~N.,  {Matsumoto} T.,   {Inutsuka} S.-i.,  2008, \mn@doi [\apj]
  {10.1086/591074}, \href
  {https://ui.adsabs.harvard.edu/abs/2008ApJ...685..690M} {685, 690}

\bibitem[\protect\citeauthoryear{Magg, Hartwig, Agarwal, Frebel, Glover,
  Griffen  \& Klessen}{Magg et~al.}{2017}]{magg17}
Magg M.,  Hartwig T.,  Agarwal B.,  Frebel A.,  Glover S.~C.,  Griffen B.~F.,
  Klessen R.~S.,  2017, Monthly Notices of the Royal Astronomical Society, 473,
  5308

\bibitem[\protect\citeauthoryear{{Marigo}, {Girardi}, {Bressan}, {Groenewegen},
  {Silva}  \& {Granato}}{{Marigo} et~al.}{2008}]{marigo08}
{Marigo} P.,  {Girardi} L.,  {Bressan} A.,  {Groenewegen} M.~A.~T.,  {Silva}
  L.,   {Granato} G.~L.,  2008, \mn@doi [\aap] {10.1051/0004-6361:20078467},
  \href {https://ui.adsabs.harvard.edu/abs/2008A&A...482..883M} {482, 883}

\bibitem[\protect\citeauthoryear{Matteucci, Panagia, Pipino, Mannucci, Recchi
  \& Della~Valle}{Matteucci et~al.}{2006}]{matteucci06}
Matteucci F.,  Panagia N.,  Pipino A.,  Mannucci F.,  Recchi S.,   Della~Valle
  M.,  2006, Monthly Notices of the Royal Astronomical Society, 372, 265

\bibitem[\protect\citeauthoryear{McConnachie}{McConnachie}{2012}]{Mccon12}
McConnachie A.~W.,  2012, The Astronomical Journal, 144, 4

\bibitem[\protect\citeauthoryear{McWilliam}{McWilliam}{1998}]{mcwilliam98}
McWilliam A.,  1998, The Astronomical Journal, 115, 1640

\bibitem[\protect\citeauthoryear{Norris, Yong, Gilmore  \& Wyse}{Norris
  et~al.}{2010}]{norris10}
Norris J.~E.,  Yong D.,  Gilmore G.,   Wyse R.~F.,  2010, The Astrophysical
  Journal, 711, 350

\bibitem[\protect\citeauthoryear{O'Donnell}{O'Donnell}{1994}]{odonnel94}
O'Donnell J.~E.,  1994, The Astrophysical Journal, 422, 158

\bibitem[\protect\citeauthoryear{{O'Shea}}{{O'Shea}}{2007}]{oshea07}
{O'Shea} B.~W.,  2007, in American Astronomical Society Meeting Abstracts. p.
  79.01

\bibitem[\protect\citeauthoryear{{Oey}}{{Oey}}{2003}]{oey}
{Oey} M.~S.,  2003, \mn@doi [\mnras] {10.1046/j.1365-8711.2003.06228.x}, \href
  {https://ui.adsabs.harvard.edu/abs/2003MNRAS.339..849O} {339, 849}

\bibitem[\protect\citeauthoryear{{Omukai} \& {Palla}}{{Omukai} \&
  {Palla}}{2001}]{omukai01}
{Omukai} K.,  {Palla} F.,  2001, \mn@doi [\apjl] {10.1086/324410}, \href
  {https://ui.adsabs.harvard.edu/abs/2001ApJ...561L..55O} {561, L55}

\bibitem[\protect\citeauthoryear{{Omukai}, {Tsuribe}, {Schneider}  \&
  {Ferrara}}{{Omukai} et~al.}{2005}]{omukai05}
{Omukai} K.,  {Tsuribe} T.,  {Schneider} R.,   {Ferrara} A.,  2005, \mn@doi
  [The Astrophysical Journal] {10.1086/429955}, \href
  {https://ui.adsabs.harvard.edu/abs/2005ApJ...626..627O} {626, 627}

\bibitem[\protect\citeauthoryear{{Pakhomov}, {Mashonkina}, {Sitnova}  \&
  {Jablonka}}{{Pakhomov} et~al.}{2019}]{Pakahmow19}
{Pakhomov} Y.~V.,  {Mashonkina} L.~I.,  {Sitnova} T.~M.,   {Jablonka} P.,
  2019, \mn@doi [Astronomy Letters] {10.1134/S1063773719050050}, \href
  {https://ui.adsabs.harvard.edu/abs/2019AstL...45..259P} {45, 259}

\bibitem[\protect\citeauthoryear{Pastorelli et~al.,}{Pastorelli
  et~al.}{2019}]{pastorelli2019}
Pastorelli G.,  et~al., 2019, Monthly Notices of the Royal Astronomical
  Society, 485, 5666

\bibitem[\protect\citeauthoryear{{Planck Collaboration} et~al.,}{{Planck
  Collaboration} et~al.}{2018}]{plank18}
{Planck Collaboration} et~al., 2018, arXiv e-prints, \href
  {https://ui.adsabs.harvard.edu/abs/2018arXiv180706209P} {p. arXiv:1807.06209}

\bibitem[\protect\citeauthoryear{Raiteri, Villata  \& Navarro}{Raiteri
  et~al.}{1996}]{Raiteri96}
Raiteri C.,  Villata M.,   Navarro J.,  1996, Astronomy and Astrophysics, 315,
  105

\bibitem[\protect\citeauthoryear{Reichert, Hansen, Hanke, Sk{\'u}lad{\'o}ttir,
  Arcones  \& Grebel}{Reichert et~al.}{2020}]{reichert20}
Reichert M.,  Hansen C.~J.,  Hanke M.,  Sk{\'u}lad{\'o}ttir {\'A}.,  Arcones
  A.,   Grebel E.~K.,  2020, Astronomy \& Astrophysics, 641, A127

\bibitem[\protect\citeauthoryear{{Ricotti} \& {Gnedin}}{{Ricotti} \&
  {Gnedin}}{2005}]{Ricotti2005}
{Ricotti} M.,  {Gnedin} N.~Y.,  2005, \mn@doi [\apj] {10.1086/431415}, \href
  {http://adsabs.harvard.edu/abs/2005ApJ...629..259R} {629, 259}

\bibitem[\protect\citeauthoryear{Romano, Bellazzini, Starkenburg  \&
  Leaman}{Romano et~al.}{2014}]{romano14}
Romano D.,  Bellazzini M.,  Starkenburg E.,   Leaman R.,  2014, Monthly Notices
  of the Royal Astronomical Society, 446, 4220

\bibitem[\protect\citeauthoryear{{Safarzadeh}, {Ji}, {Dooley}, {Frebel},
  {Scannapieco}, {G{\'o}mez}  \& {O'Shea}}{{Safarzadeh}
  et~al.}{2018}]{safarz18}
{Safarzadeh} M.,  {Ji} A.~P.,  {Dooley} G.~A.,  {Frebel} A.,  {Scannapieco} E.,
   {G{\'o}mez} F.~A.,   {O'Shea} B.~W.,  2018, \mn@doi [\mnras]
  {10.1093/mnras/sty595}, \href
  {https://ui.adsabs.harvard.edu/abs/2018MNRAS.476.5006S} {476, 5006}

\bibitem[\protect\citeauthoryear{Salvadori \& Ferrara}{Salvadori \&
  Ferrara}{2009}]{SS09}
Salvadori S.,  Ferrara A.,  2009, Monthly Notices of the Royal Astronomical
  Society: Letters, 395, L6

\bibitem[\protect\citeauthoryear{{Salvadori} \& {Ferrara}}{{Salvadori} \&
  {Ferrara}}{2012}]{SF12}
{Salvadori} S.,  {Ferrara} A.,  2012, \mn@doi [\mnras]
  {10.1111/j.1745-3933.2011.01200.x}, \href
  {https://ui.adsabs.harvard.edu/abs/2012MNRAS.421L..29S} {421, L29}

\bibitem[\protect\citeauthoryear{Salvadori, Schneider  \& Ferrara}{Salvadori
  et~al.}{2007}]{SS07}
Salvadori S.,  Schneider R.,   Ferrara A.,  2007, Monthly Notices of the Royal
  Astronomical Society, 381, 647

\bibitem[\protect\citeauthoryear{Salvadori, Ferrara  \& Schneider}{Salvadori
  et~al.}{2008}]{SS08}
Salvadori S.,  Ferrara A.,   Schneider R.,  2008, Monthly Notices of the Royal
  Astronomical Society, 386, 348

\bibitem[\protect\citeauthoryear{{Salvadori}, {Ferrara}, {Schneider},
  {Scannapieco}  \& {Kawata}}{{Salvadori} et~al.}{2010}]{SS10}
{Salvadori} S.,  {Ferrara} A.,  {Schneider} R.,  {Scannapieco} E.,   {Kawata}
  D.,  2010, \mn@doi [\mnras] {10.1111/j.1745-3933.2009.00772.x}, \href
  {https://ui.adsabs.harvard.edu/abs/2010MNRAS.401L...5S} {401, L5}

\bibitem[\protect\citeauthoryear{Salvadori, Skuladottir  \& Tolstoy}{Salvadori
  et~al.}{2015}]{SS15}
Salvadori S.,  Skuladottir A.,   Tolstoy E.,  2015, Monthly Notices of the
  Royal Astronomical Society, 454, 1320

\bibitem[\protect\citeauthoryear{Salvadori, Bonifacio, Caffau, Korotin,
  Andreevsky, Spite  \& Sk{\'u}lad{\'o}ttir}{Salvadori et~al.}{2019}]{SS19}
Salvadori S.,  Bonifacio P.,  Caffau E.,  Korotin S.,  Andreevsky S.,  Spite
  M.,   Sk{\'u}lad{\'o}ttir {\'A}.,  2019, Monthly Notices of the Royal
  Astronomical Society, 487, 4261

\bibitem[\protect\citeauthoryear{Schaerer \& Pell{\'o}}{Schaerer \&
  Pell{\'o}}{2002}]{Schaerer02}
Schaerer D.,  Pell{\'o} R.,  2002, in , Scientific Drivers for ESO Future
  VLT/VLTI Instrumentation.
Springer, pp 48--53

\bibitem[\protect\citeauthoryear{{Schneider}, {Ferrara}, {Salvaterra}, {Omukai}
   \& {Bromm}}{{Schneider} et~al.}{2003}]{Schneider2003}
{Schneider} R.,  {Ferrara} A.,  {Salvaterra} R.,  {Omukai} K.,   {Bromm} V.,
  2003, \mn@doi [\nat] {10.1038/nature01579}, \href
  {https://ui.adsabs.harvard.edu/abs/2003Natur.422..869S} {422, 869}

\bibitem[\protect\citeauthoryear{{Sestito} et~al.,}{{Sestito}
  et~al.}{2019}]{sestito19}
{Sestito} F.,  et~al., 2019, \mn@doi [\mnras] {10.1093/mnras/stz043}, \href
  {https://ui.adsabs.harvard.edu/abs/2019MNRAS.484.2166S} {484, 2166}

\bibitem[\protect\citeauthoryear{Simon}{Simon}{2019}]{Simon19}
Simon J.~D.,  2019, Monthly Notices of the Royal Astronomical Society

\bibitem[\protect\citeauthoryear{Sk{\'u}lad{\'o}ttir, Tolstoy, Salvadori, Hill,
  Pettini, Shetrone  \& Starkenburg}{Sk{\'u}lad{\'o}ttir et~al.}{2015}]{asa15}
Sk{\'u}lad{\'o}ttir {\'A}.,  Tolstoy E.,  Salvadori S.,  Hill V.,  Pettini M.,
  Shetrone M.~D.,   Starkenburg E.,  2015, Astronomy \& Astrophysics, 574, A129

\bibitem[\protect\citeauthoryear{{Smith}, {Clark}, {Glover}  \&
  {Klessen}}{{Smith} et~al.}{2010}]{smith10}
{Smith} R.~J.,  {Clark} P.~C.,  {Glover} S.,   {Klessen} R.~S.,  2010, in
  {Whalen} D.~J.,  {Bromm} V.,   {Yoshida} N.,  eds,  American Institute of
  Physics Conference Series Vol. 1294, First Stars and Galaxies: Challenges for
  the Next Decade. pp 285--286, \mn@doi{10.1063/1.3518879}

\bibitem[\protect\citeauthoryear{{Stacy} \& {Bromm}}{{Stacy} \&
  {Bromm}}{2014}]{stacybromm14}
{Stacy} A.,  {Bromm} V.,  2014, \mn@doi [\apj] {10.1088/0004-637X/785/1/73},
  \href {https://ui.adsabs.harvard.edu/abs/2014ApJ...785...73S} {785, 73}

\bibitem[\protect\citeauthoryear{Stacy, Greif, Klessen, Bromm  \& Loeb}{Stacy
  et~al.}{2013}]{stacy13}
Stacy A.,  Greif T.~H.,  Klessen R.~S.,  Bromm V.,   Loeb A.,  2013, Monthly
  Notices of the Royal Astronomical Society, 431, 1470

\bibitem[\protect\citeauthoryear{{Stacy}, {Bromm}  \& {Lee}}{{Stacy}
  et~al.}{2016}]{stacy16}
{Stacy} A.,  {Bromm} V.,   {Lee} A.~T.,  2016, \mn@doi [\mnras]
  {10.1093/mnras/stw1728}, \href
  {https://ui.adsabs.harvard.edu/abs/2016MNRAS.462.1307S} {462, 1307}

\bibitem[\protect\citeauthoryear{{Starkenburg} et~al.,}{{Starkenburg}
  et~al.}{2018}]{Starkenburg18}
{Starkenburg} E.,  et~al., 2018, \mn@doi [\mnras] {10.1093/mnras/sty2276},
  \href {https://ui.adsabs.harvard.edu/abs/2018MNRAS.481.3838S} {481, 3838}

\bibitem[\protect\citeauthoryear{{Susa}}{{Susa}}{2019}]{susa19}
{Susa} H.,  2019, \mn@doi [The Astrophysical Journal]
  {10.3847/1538-4357/ab1b6f}, \href
  {https://ui.adsabs.harvard.edu/abs/2019ApJ...877...99S} {877, 99}

\bibitem[\protect\citeauthoryear{{Tan} \& {McKee}}{{Tan} \&
  {McKee}}{2004}]{tan04}
{Tan} J.~C.,  {McKee} C.~F.,  2004, \mn@doi [\apj] {10.1086/381490}, \href
  {https://ui.adsabs.harvard.edu/abs/2004ApJ...603..383T} {603, 383}

\bibitem[\protect\citeauthoryear{{Tarumi}, {Hartwig}  \& {Magg}}{{Tarumi}
  et~al.}{2020}]{Tarumi20}
{Tarumi} Y.,  {Hartwig} T.,   {Magg} M.,  2020, \mn@doi [\apj]
  {10.3847/1538-4357/ab960d}, \href
  {https://ui.adsabs.harvard.edu/abs/2020ApJ...897...58T} {897, 58}

\bibitem[\protect\citeauthoryear{Tolstoy, Hill  \& Tosi}{Tolstoy
  et~al.}{2009}]{tolstoy09}
Tolstoy E.,  Hill V.,   Tosi M.,  2009, Annual Review of Astronomy and
  Astrophysics, 47, 371

\bibitem[\protect\citeauthoryear{{Tumlinson}}{{Tumlinson}}{2007}]{tum07}
{Tumlinson} J.,  2007, \mn@doi [\apj] {10.1086/519917}, \href
  {https://ui.adsabs.harvard.edu/abs/2007ApJ...665.1361T} {665, 1361}

\bibitem[\protect\citeauthoryear{{Turk}}{{Turk}}{2009}]{turk09}
{Turk} M.~J.,  2009, PhD thesis, Stanford University

\bibitem[\protect\citeauthoryear{Venn, Irwin, Shetrone, Tout, Hill  \&
  Tolstoy}{Venn et~al.}{2004}]{venn04}
Venn K.~A.,  Irwin M.,  Shetrone M.~D.,  Tout C.~A.,  Hill V.,   Tolstoy E.,
  2004, The Astronomical Journal, 128, 1177

\bibitem[\protect\citeauthoryear{Vincenzo, Matteucci, Vattakunnel  \&
  Lanfranchi}{Vincenzo et~al.}{2014}]{vincenzo14}
Vincenzo F.,  Matteucci F.,  Vattakunnel S.,   Lanfranchi G.~A.,  2014, Monthly
  Notices of the Royal Astronomical Society, 441, 2815

\bibitem[\protect\citeauthoryear{{Webster}, {Bland-Hawthorn}  \&
  {Sutherland}}{{Webster} et~al.}{2015}]{webster15}
{Webster} D.,  {Bland-Hawthorn} J.,   {Sutherland} R.,  2015, \mn@doi [\apjl]
  {10.1088/2041-8205/799/2/L21}, \href
  {https://ui.adsabs.harvard.edu/abs/2015ApJ...799L..21W} {799, L21}

\bibitem[\protect\citeauthoryear{Weidner \& Kroupa}{Weidner \&
  Kroupa}{2006}]{weidner06}
Weidner C.,  Kroupa P.,  2006, Monthly Notices of the Royal Astronomical
  Society, 365, 1333

\bibitem[\protect\citeauthoryear{Weidner, Kroupa, Pflamm-Altenburg  \&
  Vazdekis}{Weidner et~al.}{2013}]{weidner13}
Weidner C.,  Kroupa P.,  Pflamm-Altenburg J.,   Vazdekis A.,  2013, Monthly
  Notices of the Royal Astronomical Society, 436, 3309

\bibitem[\protect\citeauthoryear{{Wise}, {Abel}, {Turk}, {Norman}  \&
  {Smith}}{{Wise} et~al.}{2012}]{Wise12}
{Wise} J.~H.,  {Abel} T.,  {Turk} M.~J.,  {Norman} M.~L.,   {Smith} B.~D.,
  2012, \mn@doi [\mnras] {10.1111/j.1365-2966.2012.21809.x}, \href
  {https://ui.adsabs.harvard.edu/abs/2012MNRAS.427..311W} {427, 311}

\bibitem[\protect\citeauthoryear{{Wollenberg}, {Glover}, {Clark}  \&
  {Klessen}}{{Wollenberg} et~al.}{2020}]{wollenberg20}
{Wollenberg} K. M.~J.,  {Glover} S. C.~O.,  {Clark} P.~C.,   {Klessen} R.~S.,
  2020, \mn@doi [\mnras] {10.1093/mnras/staa289}, \href
  {https://ui.adsabs.harvard.edu/abs/2020MNRAS.494.1871W} {494, 1871}

\bibitem[\protect\citeauthoryear{{Woosley} \& {Weaver}}{{Woosley} \&
  {Weaver}}{1995}]{Woo95}
{Woosley} S.~E.,  {Weaver} T.~A.,  1995, \mn@doi [\apjs] {10.1086/192237},
  \href {https://ui.adsabs.harvard.edu/abs/1995ApJS..101..181W} {101, 181}

\bibitem[\protect\citeauthoryear{Yoshida, Abel, Hernquist  \& Sugiyama}{Yoshida
  et~al.}{2003}]{yoshida03}
Yoshida N.,  Abel T.,  Hernquist L.,   Sugiyama N.,  2003, The Astrophysical
  Journal, 592, 645

\bibitem[\protect\citeauthoryear{{Yoshida}, {Omukai}, {Hernquist}  \&
  {Abel}}{{Yoshida} et~al.}{2006}]{yoshida06}
{Yoshida} N.,  {Omukai} K.,  {Hernquist} L.,   {Abel} T.,  2006, \mn@doi [\apj]
  {10.1086/507978}, \href
  {https://ui.adsabs.harvard.edu/abs/2006ApJ...652....6Y} {652, 6}

\bibitem[\protect\citeauthoryear{de Bennassuti, Salvadori, Schneider, Valiante
  \& Omukai}{de~Bennassuti et~al.}{2017}]{debennessuti17}
de Bennassuti M.,  Salvadori S.,  Schneider R.,  Valiante R.,   Omukai K.,
  2017, Monthly Notices of the Royal Astronomical Society, pp 926--940

\bibitem[\protect\citeauthoryear{de Jong et~al.,}{de~Jong
  et~al.}{2019}]{dejong4MOST}
de Jong R.~S.,  et~al., 2019, arXiv preprint arXiv:1903.02464

\makeatother
\end{thebibliography}




\appendix
\section{Hercules, Leo IV, and Eridanus II}
\begin{figure}
\includegraphics[width=0.9\columnwidth]{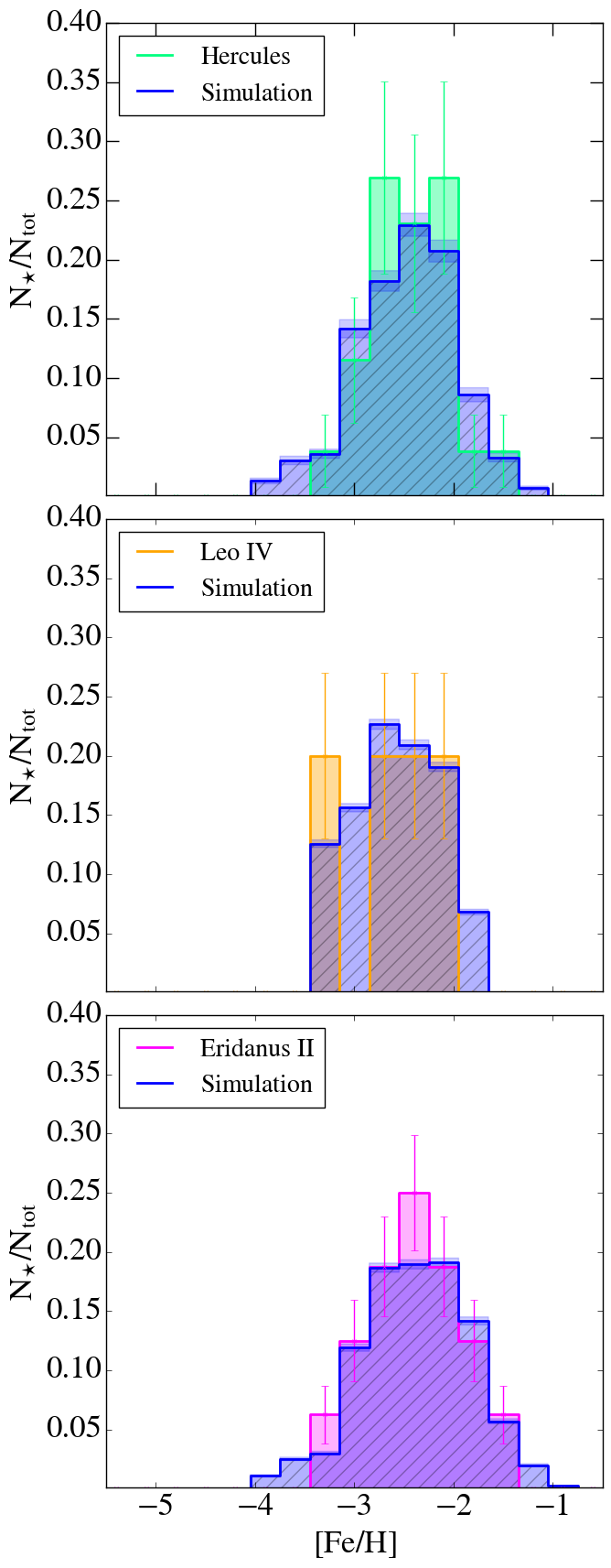}
\caption{The comparison between the observed and the simulated (blue) MDFs for UDFs in the $99\%$ of confidence level $\chi^2$ contour of Bo{\"o}tes~I: Hercules (lightgreen), Leo~IV (orange) and Eridanus~II (magenta). }
\label{UFDs_MDFs}
\end{figure}  
\label{MDFsUFDs}

Fig.\ref{UFDs_MDFs} shows the comparison between the observed and the simulated MDFs for those UFDs that reside in the $99\%$ confidence level $\chi^2$ contour of Bo{\"o}tes~I, i.e. Hercules, Leo~IV and Eridanus~II (see Fig.\ref{modelcalcalibration}, Sec.\ref{modelcal}). We can see that in all cases we find a good agreement between model results and observations.

\section{PopIII IMF random sampling}
\label{appendixA1}
\begin{figure*}
	\includegraphics[width=\textwidth]{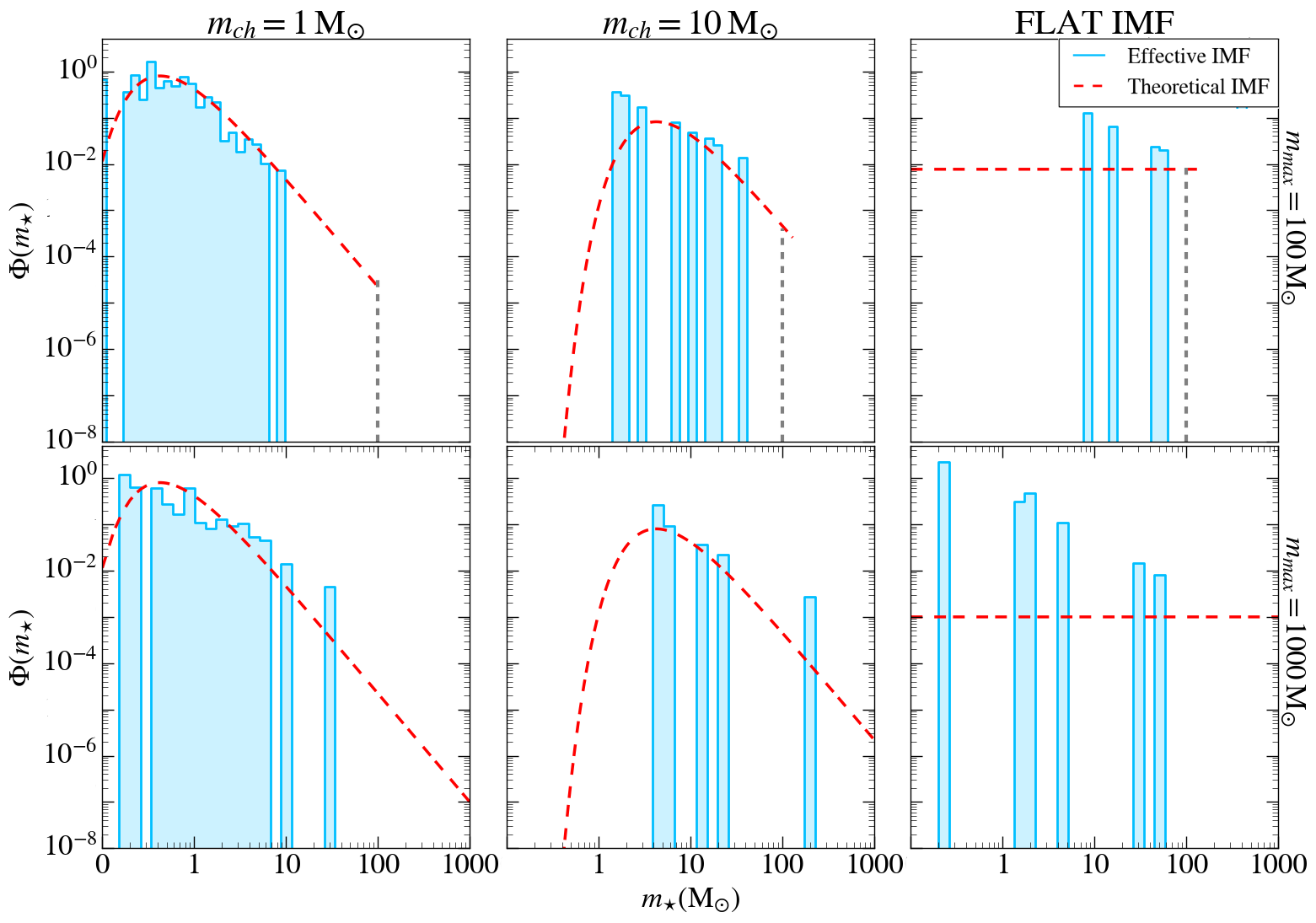}
  \caption{The comparison between the theoretical IMF (red dashed line) and the effective one (blue histograms) for different shapes and maximum mass, $m_{max}$, of the PopIII IMF. We show a Larson IMF with $m_{ch}=1\msun$ (left panels), $m_{ch}=10\msun$ (middle panels), and a flat IMF (right panels). The top (bottom) panels show results for $m_{max}=100\msun$ ($m_{max}=1000\msun$). 
 }
\label{samplingmch}
\end{figure*}

Fig.\ref{samplingmch} shows the comparison between the theoretical PopIII IMF (see Fig.\ref{imfs}) and the effective one obtained with the random sampling procedure, when we assume to form a total stellar mass of $\rm \approx 200 \msun$. This the typical mass formed in a single burst during the early evolutionary phases of Bo{\"o}tes~I, i.e. when $\Psi \approx 2 \cdot 10^{-4} \msun \rm yr^{-1}$ (Fig.~\ref{SFR}). We see that the theoretical PopIII IMF is poorly sampled independent of its shape, the $m_{ch}$, and the $m_{max}$. Furthermore, while for $m_{ch}=1 \msun$ we still have a good sampling of the PopIII IMF at the lowest mass-end, $m_{\star} < 1 \msun$, when the peak of the IMF is shifted towards higher masses ($m_{ch} = 10 \msun $) or there is not a preferential mass scale, i.e. the IMF is flat, the lowest mass-end becomes less and less populated with respect to the theoretical IMF. This implies that the number of surviving PopIII stars is much lower than expected. These results are even more extreme for $m_{max}=1000\msun$ (bottom panels).\\

\section{IMF sampling: Bo{\"o}tes~I evolution}
\label{appendixB}

\begin{figure*}
	\includegraphics[width=0.47\textwidth]{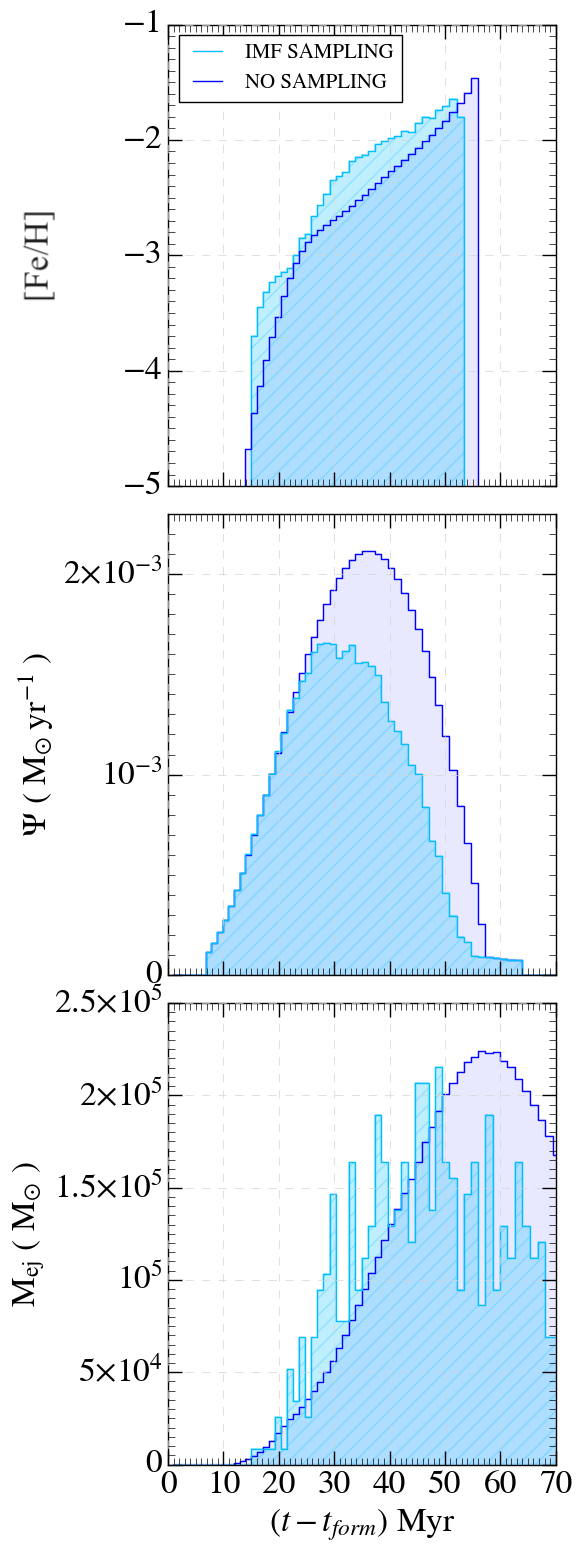}
  \caption{The evolution of iron-abundance (top panel), star-formation rate (medium panel) and the ejected gas mass (bottom panel) for the model with (light blue) and without (dark blue) IMF random sampling.
}
\label{conf}
\end{figure*}

In Fig.~\ref{conf} we show the evolution of the ISM iron abundance, $[\rm Fe/H]$, the star formation rate, $\Psi$, and the ejected gas mass, $\rm M_{ej}$, of Bo{\"o}tes~I when we consider models with and without the random sampling of the IMF.
In the bottom panel of Fig.\ref{conf} we see that $[\rm Fe/H]$ in the random sampling model starts with a time-delay of $\sim 2$Myr. This is because the more massive SNe, $m_{\star} = 40 \msun$, that explode first are not produced in this early burst of star formation, which has $\Psi \approx 10^{-4}\msun \rm yr^{-1}$ (middle panel). Still, as soon as low-mass SNe stars to explode, $[\rm Fe/H]$ in the IMF sampling model rapidly grows, exceeding the value of the model without sampling (top) in spite of the equal rate of star formation (middle panel). This is because the random sampling of the IMF provide us with a finite number of SNe instead of a fraction. Although delayed, therefore, the effect of feedback from SNe is stronger on both the ISM enrichment (top) and the mass of gas ejected (bottom panel). When $(t-t_{form})\approx 25$~Myr, i.e. at the peak of the star formation for the IMF sampling model, we find that [Fe/H]$\approx -2.6$, which corresponds to the peak of the MDF (Fig.~\ref{MDFconfronto}). In the model without IMF sampling the star formation rate reaches higher values but this maximum appears at later times, $(t-t_{form})\approx 35$~Myr. Yet, because of the lower metal enrichment [Fe/H]$\approx -2.7$. This explains why the MDF is higher but shifted towards lower [Fe/H] values in the model without IMF sampling (Fig.~\ref{MDFconfronto}). 

\bsp	
\label{lastpage}

\end{document}